\documentclass[12pt]{article}
\usepackage{graphicx}
\usepackage{mathptmx,mathrsfs}
\usepackage{amsfonts,amsmath,amssymb,amsthm}
\usepackage{indentfirst}
\usepackage{cite}
\usepackage{hyperref}
\usepackage{enumitem}

\numberwithin{equation}{section}
\newtheorem{theorem}{Theorem}[section]

\theoremstyle{definition}

\begin{document}

\title{The Epstein-Glaser causal approach to the Light-Front QED$_{4}$. II: Vacuum Polarization tensor}
	\author{R. Bufalo$^{1,2}$\thanks{%
rbufalo@ift.unesp.br}~, B.M. Pimentel$^{2}$\thanks{%
pimentel@ift.unesp.br}~, D.E. Soto$^{2}$\thanks{%
danielsb@ift.unesp.br}~ \\
\textit{{$^{1}${\small Department of Physics, University of Helsinki, P.O. Box 64}}}\\
\textit{\small FI-00014 Helsinki, Finland}\\
\textit{{$^{2}${\small Instituto de F\'{\i}sica Te\'orica (IFT), UNESP,  S\~ao Paulo State University}}} \\
\textit{\small Rua Dr. Bento Teobaldo Ferraz 271, Bloco II Barra Funda, CEP
01140-070 S\~ao Paulo, SP, Brazil}\\
}
\maketitle
\date{}

\begin{abstract}
In this work we show how to construct the one-loop vacuum polarization for light-front QED$_{4}$ in the framework of the perturbative causal theory. Usually, in the canonical approach, it is considered for the fermionic propagator the so-called \emph{instantaneous} term, but it is known in the literature that this term is controversial because it can be omitted by computational reasons; for instance, by compensation or vanishing by dimensional regularization. In this work we propose a solution to this paradox. First, in the Epstein-Glaser causal theory, it is shown that the fermionic propagator does
not have instantaneous term, and with this propagator we calculate the one-loop vacuum polarization, from this calculation it follows the same result as those obtained by the standard approach, but without reclaiming any extra assumptions. Moreover, since the perturbative causal theory is defined in the distributional framework, we can also show the reason behind we obtaining the same result whether we consider or not the instantaneous fermionic propagator term.
\end{abstract}

\newpage

\section{Introduction}

Perturbative Quantum Electrodynamics (QED$_4$) is a gauge theory that presents a remarkable computational success. For instance, one may cite its impressive accuracy with the measurement of the anomalous magnetic moment of the electron and the muon \cite{ref45}. However, one may even wonder if the physical dynamics of QED$_4$ is preserved, or how it changes,
if it is defined in the light-front coordinates. It is a well-known fact that in such form of dynamics there are non established issues concerning the importance of the \textit{instantaneous} terms in the diagrams (the problem is the interpretation and true meaning of such contributions) in order to realize whether or not they are physically relevant though not propagating any information.

One may say that the most natural approach for massless fields, such as the electromagnetic field, is given in the light-front dynamics. This approach was proposed initially by Dirac \cite{3} in 1949, he showed different choices of the time evolution parameter \footnote{Given by the light-front $\{x^{\pm} \sim x^{0}\pm x^{3}\}$ or the usual instant-form $\{x^{0}\}$, and they are not related by a Lorentz transformation of coordinates.} Moreover, the light-front quantization \cite{ref35} is very appealing and simple in the sense that it is rather economical in displaying the relevant degrees of freedom of a given theory; and, thus, the discussion of the physical Hilbert space is more tractable, and the physical vacuum state is trivial \cite{ref30}. This fact leads to interesting analysis of nonperturbative effects in the context of QCD$_4$ \cite{ref37}. We may also cite studies upon the exact solution of two-dimensional BF and Yang-Mills theories in the light-front \cite{ref25}.

The initial attempts undertaken in attaining the canonical quantization of the light-front QED$_4$ in the light-cone gauge $A_{-}=0$ have been known for almost forty years by now \cite{ref5,ref6,ref7}. However, some difficulties and inconsistencies remain
present in this approach, some of these problems were associated with the gauge choice: Feynman amplitudes at the one-loop level exhibited double-pole singularities \cite{ref33}. This pathological behavior has been ascribed to the Principal Value (PV) prescription employed to the treatment of the poles $\left( k.n\right) ^{-1}$ in the gauge boson propagator \cite{ref8}.

It was realized later that, in order to handle to these poles, it was need to prescribe methods to circumvent the pathology. For instance, we have that, in the electromagnetic doubly transverse gauge arises poles of the form%
\begin{equation}
g\left(k;n\right)=\frac{1}{\left( k^{+}\right) ^{n}},  \label{eq 0.0}
\end{equation}%
where $n=1,2$. Usually they are named as "spurious" poles. To handle to this problem, several different prescriptions were
proposed in the light-front form. Among the most known in the literature, we may cite the Mandelstam-Leibbrant prescription \cite{ref10} and Pimentel-Suzuki prescription \cite{ref29}. The latter has as the basic premise the fact that the propagator as a whole must be \textit{causal} in order to treat the light-cone pole (also the higher-order poles). This has showed, that mathematics only does not suffice for such a task. Moreover, the above mentioned prescriptions were designed in order to ensure that the location of the poles in the $k^{0}$-plane -- located in the second and fourth quadrants -- would not hinder Wick rotation nor spoil power-counting. Recently the authors have analysed the free fields of QED$_4$ \cite{ref28}, discussing the analytic representation of the propagators and commutators, and also showing how the causal method of Epstein-Glaser \cite{ref17,ref19} may be used in order to handle poles as those in the Eq.\eqref{eq 0.0} for $n=1,2,3,\ldots$ without referring to any particular prescription, recovering the results of both aforementioned prescription as particular cases. Actually, there are other interesting studies in this direction \cite{ref40}.

Returning to the perturbative studies of QED$_4$, there are many interesting issues being discussed in the literature lately. For instance, the use of coherent states to deal with the infrared divergences in light-front QED$_4$ \cite{ref31}
and the study of the equivalence between the covariant and light-front QED$_4$ \cite{ref39}. Despite the amount of substantial activity in the area, there are studies of some issues in the QED$_4$ that have not been fully discussed in all aspects,
and certainly they deserve a new detailed treatment and interpretation. As mentioned before, there is the controversial \emph{instantaneous} term in the fermionic propagator, which is derived in the canonical theory but omitted in the practice by not so clear arguments: by compensation reasons \cite{ref12}, or vanishing in dimensional regularization calculation \cite{ref14}. And it is precisely there, handling with the interpretation and contribution of this instantaneous term, where we will focus our discussion in this paper.

We believe that the origin of these misleading results in the light-front QED$_4$ may be attributed to the nonrigorous mathematical aspects in the general approaches and as well as by the fact that general properties, such as causality, are not carefully implemented. Besides, one may also emphasize the lack of consistence and rigorous in dealing naively with a field theory by only making a change of variables to light-front coordinates when performing diagram Feynman integrals, or even by constructing from the very beginning the field theory defined in the light-front coordinates. Recalling that in the perturbative study of QED$_4$ we have that the series expansion of the \emph{S-Matrix} takes the form%
\begin{equation}
S=1+\sum\limits_{n=1}^{\infty }\frac{1}{n!}\int
dx_{1}dx_{2}...dx_{n}S_{n}\left( x_{1},x_{2},...,x_{n}\right) \colon
\prod\limits_{j=1}^{f_{g}}\bar{\psi}\left( x_{k_{j}}\right)
\prod\limits_{j=1}^{f_{g}}\psi \left( x_{n_{j}}\right) \colon
:\prod\limits_{j=1}^{l_{g}}A_{\mu _{m_{j}}}\left( x_{m_{j}}\right) :,
\end{equation}%
where in the last part $\colon\colon$ stands for the normal product of free fermionic fields and electromagnetic field, respectively. The coefficient function $S_{n}\left(x_{1},x_{2},...,x_{n}\right) $ is expressed either in terms of the fermionic and/or
electromagnetic propagators, and we may obtain them explicitly by evaluating the temporal ordering products. As it was pointed out by Bogoliubov and Parasiuk \cite{ref4}, it is a long-term mathematical problem the fact that the products of
Heaviside functions and $\delta $-Dirac distributions like: $\theta \left(x\right) \delta \left( x\right) $, are recognized as being the origin of the so called ultraviolet (UV) divergences. In the standard canonical theory the finiteness of the results, which are in agreement with experiments, is achieved only after a series of steps, starting by the regularization of divergent integrals, and, subsequently, the absorption of these regularized infinities into the physical quantities
(mass, charge, and etc), this is the well-known procedure of renormalization.

In an unorthodox line of development, one looks for an approach defined in such a way that this shall be mathematically consistent, which works with well-defined products. This points directly towards the use of the distribution theory \cite{ref26,ref27} to deal with these intriguing quantities. For instance, in the construction of the coefficient functions $S_{n}$, the vacuum expectation value of normally product of fields plays an important role. A theory defined in a distributional framework, where this product of fields (Wightman functions) plays a central role is given by the pioneer Wightman formalism \cite{ref15}, this is an axiomatic quantum field theory which considers as postulates the following set of physical requirements: the quantum mechanical framework, relativistic invariance, existence and uniqueness of the vacuum, fields as an operator-valued distribution, spectral condition and locality \cite{ref15}.

Many efforts have been done in the development of a mathematically rigorous field theory, and in 1973 Epstein and Glaser \cite{ref17} proposed the \emph{perturbative causal theory}, which is an axiomatic perturbative theory for the \emph{S-Matrix} that considers the following postulates: \emph{causality}, \emph{relativistic invariance} and \emph{asymptotic
conditions}. \footnote{The ideas of this formulation were first introduced by Heisenberg \cite{1}.} This method was formulated in order to give a mathematical rigorous treatment for the ultraviolet divergences in quantum field theory. In such framework such divergences do not appear anywhere in the calculations due to the correct splitting of the causal distributions into its advanced and retarded parts. Later, this approach was implemented in the practical momentum space framework and applied in several field models, for instance, to QED$_4$ and QCD$_4$ \cite{ref19}, QED$_3$ \cite{ref85}, Gauged Thirring model \cite{ref87}, and the DKP theory \cite{ref13}, in the usual instant-form.

In Ref.\cite{ref28} we have focused in applying the causal method in the study of free fields in the light-front
and also to accomplish a solution to the problem of the spurious poles of the electromagnetic propagator. Hence,
this leads to the thought that through the Epstein-Glaser's causal method we can handle the problem of the instantaneous
part of the fermionic propagator in a suitable and proper fashion. For this purpose, in Sect.\ref{sec:1} we start by
reviewing the general properties of the Epstein-Glaser's causal method, by presenting a complete explanation of the
necessary modifications in order to implement the inductive construction of the method for the light-front framework.
Next, in Sect.\ref{sec:2}, we shall apply this approach to study the one-loop vacuum polarization of QED$_{4}$ at light-front, and, subsequently, in Sect.\ref{sec:3}, we analyze and discuss the possible modifications when we consider the instantaneous part of the fermionic propagator in the computation. In Sect.\ref{sec:4} we summarize the results, and present our final remarks and prospects.

\section{Perturbative causal theory in the light-front}

\label{sec:1}

One of the most important object in quantum electrodynamics is the scattering \emph{S-Matrix} which encodes all the information about the lepton-photon interaction processes. With the help of \emph{S-Matrix} we can calculate other basic quantities of QED, such as the \emph{Green's functions}. In the usual approach these are vacuum expectation values of time ordered product of fields and they can be calculated perturbatively by means of the Feynman rules, also regularization and renormalization procedure are required. Thus, for perturbative Light-Front QED, some basic Green's functions are the fermionic and electromagnetic (Feynman) free propagators, given by
\begin{align}
S^{F}\left( x\right)  =&\theta \left( x^{+} \right) \left\langle 0
|\psi \left( x\right) \bar{\psi}\left( 0\right) |\Omega \right\rangle
-\theta \left( -x^{+} \right) \left\langle 0 |\bar{\psi}\left( 0\right)
\psi \left( x\right) |0 \right\rangle , \\
D_{\mu \nu }^{F}\left( x\right)  =&\theta \left( x^{+} \right) \left\langle
0 |A_{\mu }\left( x\right) A_{\nu }\left( 0\right) |0
\right\rangle +\theta \left( -x^{+} \right) \left\langle 0 |A_{\nu
}\left( 0\right) A_{\mu }\left( x\right) |0 \right\rangle ,
\end{align}
respectively; $ (\psi,\bar{\psi}) $, $ A_{\mu } $ are the fermionic and electromagnetic free fields, respectively, and $ \left\vert 0\right\rangle $ is the vacuum state and $ x^{+} $ is the temporal parameter in the light-front dynamics.

The Perturbative causal method, proposed by Epstein-Glaser \cite{ref17} to quantum field theory, is an axiomatic perturbative formalism of the \emph{S-Matrix} which formulation only considers well-defined products, hence no regularization method it needed. The causal approach considers the scattering matrix, \emph{S-Matrix}, as proposed by Bogoliubov \cite{ref22}, in which it is an operator-valued functional and can be written in the following purely formal perturbative series
\begin{equation}
S\left[ g\right] =1+\sum\limits_{n=1}^{\infty }\frac{1}{n!}\int
dx_{1}dx_{2}...dx_{n}T_{n}\left( x_{1},x_{2},...,x_{n}\right) g\left(
x_{1}\right) g\left( x_{2}\right) ...g\left( x_{n}\right) ,
\label{eq 1.3}
\end{equation}%
where we can identify the symmetric $n$-point function $T_{n}$ as an operator-valued distribution and $g^{\otimes n}$ its respective test function, moreover, it is supposed to belong to the Schwartz space, $g\left( x\right)\in\mathcal{J}\left( \mathbf{M}^{4}\right) $. The test function plays the role of switching the interaction \emph{in} some region of the spacetime, thus, when the interaction is completely switched \emph{off}, $g=0$, the \emph{S-Matrix} is the identity operator: $ S\left[ 0\right] =I $, as we see from \eqref{eq 1.3}.

An advantage of the causal approach is that only free asymptotic fields acting on the Fock space (well-defined quantities) are utilized in order to construct $S\left[ g\right]$. And the building blocks $T_{n}$ are constructed via an \textit{inductive method}, which is established when we consider a few assumptions or axioms: \emph{causality}, introduced by St\"{u}ckelberg  \cite{ref16}; \emph{relativistic invariance}, introduced by Wigner \cite{2}; and finally the \emph{asymptotic conditions}, as proposed by Heisenberg \cite{1}. The causal approach has been implemented to the usual (instant form) QED by Scharf \textit{et al} \cite{ref19}, for the implementation to the light-front form we must review each step of the usual form and introduce some alterations where it makes necessary.

\subsection{General properties}

As proposed initially by N.N. Bogoliubov and collaborators \cite{ref22}, some basic physical assumptions are needed in order to construct the scattering matrix $ S=S\left[ g\right] $ with the help of the adiabatic switching. Moreover, as aforementioned the causal approach yields the \emph{S-Matrix} directly in the Fock space of well-defined free fields.

\emph{Causality}.- A physical observer must be able to localize such as order events in the spacetime. This is achieved by a parameter called "time", denoted by $\tau=\eta \left( \partial _{\tau },x\right) $; where $\eta $ is the metric tensor, while $\partial _{\tau }$ is the tangent vector of the observer world-line and $ x $ is some event. Now consider two
test function $g_{1}$ and $g_{2}$ with disjoint supports, then if the support of $g_{1}$ is earlier than the support of $g_{2}$,  $\{\forall ~ x_{1}\in \text{Supp}~\left( g_{1}\right) \}$ and $\{\forall ~ x_{2}\in \text{Supp}~\left( g_{2}\right) \}$, i.e.: $ \text{Supp}~\left( g_{1}\right) <\text{Supp}~\left( g_{2}\right)$,\footnote{A physical observer follows a time-like curve, so if we consider the light-front coordinates $\left( x^{+},x^{1},x^{2},x^{-}\right) $, then to guarantee that $x_{1}<x_{2}$ is a \textit{relativistic causal relation} is necessary that $x_{1}^{+}<x_{2}^{+}$ and $x_{1}^{-}<x_{2}^{-}$, both together. See \ref{app:A} for our basic notation.} then the \emph{S-Matrix} $S\left[g_{1}+g_{2}\right] $ satisfies
\begin{equation}
S\left[ g_{1}+g_{2}\right] =S\left[ g_{2}\right] S\left[ g_{1}\right] ,  \label{eq 1.5}
\end{equation}%
this is the causal formulation of the \emph{S-Matrix}. From this property follows that, when we replace the perturbative series \eqref{eq 1.3}, we arrive at the causal relation for the $ T_{n}$ distributions:
\begin{equation}
T_{n}\left( x_{1},...,x_{m},x_{m+1},...,x_{n}\right) =T_{m}\left(
x_{1},...,x_{m}\right) T_{n-m}\left( x_{m+1},...,x_{n}\right) ,
\label{eq 1.6}
\end{equation}%
where it holds: $\left\{ x_{1},...,x_{m}\right\} >\left\{ x_{m+1},...,x_{n}\right\} $ \footnote{Which means that $\tau _{j}>\tau _{i}$, for $j=1,\ldots ,m$ while $i=m+1,\ldots ,n$.}. From this relation one may conclude that $T_{n}$ is a \emph{causal ordering product} distribution. Moreover, one can easily realize that since the sign $>$ is understood in \textit{stricto sensu}, the distribution $T_{n}$ can not be expressed in terms of the well-known Feynman time-ordering product: $ T_{n}\left( x_{1},...,x_{n}\right) \neq \mathcal{T}\left[ T_{1}\left( x_{1}\right) \cdots T_{1}\left( x_{n}\right) \right] $, known to originate the UV divergences. \footnote{By definition: $ \mathcal{T}\left[ T_{1}\left( x_{1}\right) \cdots T_{1}\left( x_{n}\right) \right] =\sum\limits_{\pi }\theta \left( x_{\pi \left( 1\right) }^{+}-x_{\pi
\left( 2\right) }^{+}\right) \cdots \theta \left( x_{\pi \left( n-1\right) }^{+}-x_{\pi \left( n\right) }^{+}\right) T_{1}\left( x_{\pi \left( 1\right) }\right) \cdots T_{1}\left( x_{\pi \left( n\right) }\right) $.}

\emph{Asymptotic condition and interaction}.- There must be an asymptotic spacetime region where the fields are defined in terms of the free fields Fock space: $\mathcal{F}_{in}$, $\mathcal{F}_{out}$ for $\tau \rightarrow -\infty $ and $\tau \rightarrow +\infty $, respectively. The full spaces are constructed from the successive action of free field operators at the vacuum state; for instance, for QED$_4$, we have the electromagnetic and fermionic free fields: $A_{\mu }$, $\left( \psi ,\bar{\psi}\right) $. Besides, it also follows from this axiom the following reasonable assumption: at the limit $ g\rightarrow 1$, this perturbative quantum field theory has the very same first coupling perturbative term \cite{ref18}. Then for QED$_4$, the term $T_{1}$ takes the following form:
\begin{equation}
T_{1}\left( x\right) =ie\colon \bar{\psi}\left( x\right) \gamma ^{\mu }\psi
\left( x\right) \colon A_{\mu }\left( x\right) , \label{eq 0.2}
\end{equation}%
where $e$ is the coupling constant and the symbol $\colon \colon $ indicates the normal ordering product.

\emph{Relativistic invariance}.- In general $\mathcal{U}$ is a symmetry if for two observers $O$ and $O^{\prime }$, which look the same system, the transition probabilities are equal. Furthermore, in our case it follows that each observer defines its \textit{S-Matrix} as the following
\begin{equation}
S:\mathcal{F}_{in}\rightarrow \mathcal{F}_{out}, \qquad S^{\prime }:%
\mathcal{F}_{in}^{\prime }\rightarrow \mathcal{F}_{out}^{\prime }.
\end{equation}%
respectively. Now, if we consider the situation where $\mathcal{F}_{in}=\mathcal{F}_{out}=\mathcal{F}$, then the
symmetry $\mathcal{U}$ can be represented by a single operator $U$ acting at:$\mathcal{F}_{in\left( out\right) }\rightarrow \mathcal{F}_{in\left( out\right) }^{\prime }$. Thus, it follows that we can write the following (not necessarily unitary) similarity transformation
\begin{equation}
S^{\prime }=U S U^{-1}.  \label{eq 0.20}
\end{equation}
In order to discuss carefully the symmetries, the causal perturbative theory considers only the following two relativistic
invariance $U(\Lambda, a)$:

\emph{Translational invariance $x\rightarrow x^{\prime }=x+a$}: If $U\left( 1,a\right) $ is the operator which represents this symmetry on the free particle Fock space $\mathcal{F}$, then it follows from \eqref{eq 0.20} that we have the following relation
\begin{equation}
U\left( 1,a\right) S\left[ g\right] U^{-1}\left( 1,a\right) =S\left[ g_{a}%
\right], \qquad g_{a}\left( x\right) =g\left( x-a\right).
\end{equation}%
Moreover, if one replaces the formal perturbative series \eqref{eq 1.3}, it follows the following relation:
\begin{equation}
T_{n}\left( x_{1},x_{2},...,x_{n}\right) =T_{n}\left(
x_{1}-x_{n},x_{2}-x_{n},...,x_{n-1}-x_{n}\right) ,
\end{equation}%
this last form provides a great advantage when defining this distribution at momentum space.

\emph{Lorentz Invariance $x\rightarrow x^{\prime }=\Lambda x$}: Then the action of the symmetry operator $U\left( \Lambda ,0\right) $ into $S$ leads to
\begin{equation}
U\left( \Lambda ,0\right) S\left[ g\right] U^{-1}\left( \Lambda ,0\right) =S%
\left[ g_{\Lambda }\right], \qquad g_{\Lambda
}\left( x\right) =g\left( \Lambda ^{-1}x\right).
\end{equation}%
From this result we obtain that:%
\begin{equation}
U\left( \Lambda ,0\right) T_{n}\left( x_{1},x_{2},...,x_{n}\right)
U^{-1}\left( \Lambda ,0\right) =T_{n}\left( \Lambda x_{1},\Lambda
x_{2},...,\Lambda x_{n}\right) .
\end{equation}

\subsection{Inductive Construction of the S-Matrix}

The inductive method states that the n-order distribution $T_{n}(x_1,...,x_n)$ can be constructed from lower-order distributions. First we shall need to introduce some general results of the perturbative series of the inverse dispersion operator: $S^{-1}\left[g\right]$, which must fulfil the following relations%
\begin{equation}
S^{-1}\left[ g\right] S\left[ g\right] =S\left[ g\right] S^{-1}\left[ g%
\right] =1,  \label{eq 0.3}
\end{equation}%
and, analogously to $S$, in the Eq.\eqref{eq 1.3}, this inverse operator can be expressed by a formal perturbative series%
\begin{equation}
S^{-1}\left[ g\right] =1+\sum\limits_{n=1}^{\infty }\frac{1}{n!}\int
dx_{1}dx_{2}...dx_{n}\tilde{T}_{n}\left( x_{1},x_{2},...,x_{n}\right)
g\left( x_{1}\right) g\left( x_{2}\right) ...g\left( x_{n}\right) ,
\label{eq 0.4}
\end{equation}%
where the symmetric quantity $\tilde{T}_{n}\left( x_{1},...,x_{n}\right)$ is an operator-valued distribution and $%
g^{\otimes n}$ is its test function. We may compute the distributions $\tilde{T}_{n}$ by replacing the two perturbative series \eqref{eq 1.3} and \eqref{eq 0.4} into \eqref{eq 0.3}
\begin{equation}
\tilde{T}_{n}\left( X_{n}\right) =\sum_{r=1}^{n}\left( -1\right)
^{r}\sum_{P_{r}}\left[ T_{n_{1}}\left( X_{1}\right) ...T_{n_{r}}\left(
X_{r}\right) \right] ,
\end{equation}%
where $P_{r}$ stands as all partitions of the set $X_{n}=\left\{ x_{1},\ldots ,x_{n}\right\} $ into $r$ disjoint subsets nonempty: $X_{n}=\bigcup\limits_{j=1}^{r}X_{j}$, $ X_{j}\neq \emptyset $, and $\left\vert X_{j}\right\vert =n_{j}$. From this relation it follows that if we know the set $\left\{ T_{1},\ldots ,T_{n-1}\right\} $ we can determine the distribution $\tilde{T}_{n}$.\footnote{\textbf The general principles introduced previously to the distributions $T_{n}$, may be
analogously applied to the distributions $\tilde{T}_{n}$.}

Normally, the next step of the perturbation theory approach would be to use the standard formal Feynman time-ordering of $T_1$ in order to determine $T_n$, but we know now that it contains pathological ultraviolet divergences. And it is precisely that the crucial point where Epstein and Glaser proceeded more carefully and introduced the following well-defined distributional product: $ T_{p}\left( X\right) \tilde{T}_{n-p}\left( Y\right) $ and $ \tilde{T}_{p}\left( X\right) T_{n-p}\left( Y\right) $, with $X\cap Y=\emptyset $. Hence, we can define the following intermediate $n$-point distributions:%
\begin{align}
A_{n}^{\prime }\left( x_{1},...,x_{n}\right) &\equiv \sum_{P_{2}}\tilde{T}%
_{n_{1}}\left( X\right) T_{n-n_{1}}\left( Y,x_{n}\right), \label{eq 0.12}\\
R_{n}^{\prime }\left( x_{1},...,x_{n}\right) &\equiv
\sum_{P_{2}}T_{n-n_{1}}\left( Y,x_{n}\right) \tilde{T}_{n_{1}}\left(X\right) , \label{eq 0.13}
\end{align}%
where $P_{2}$ are all partitions of the set $\left\{ x_{1},...,x_{n-1}\right\} =X \bigcup Y$ into the disjoint sets $X$ and $Y$ in such a way that $\left\vert X\right\vert =n_{1}\geq 1$ and $\left\vert Y\right\vert \leq n-2$. From these distributions it follows an important property about the causal relations between the set of points $\left( x_{1},...,x_{n}\right) $, this property may be given in the form of the following theorem:

\begin{theorem}
.- Given the set $Y=P\cup Q$ such that $P\neq \emptyset $, $P\cap Q=\emptyset $, $\left\vert Y\right\vert =n_{1}\leq n-1$, and the point $x_{n}\notin Y$, then, it follows two cases:

If $\left\{ Q,x_{n}\right\} >P$, $\left\vert Q\right\vert =n_{2}$, we have that%
\begin{equation}
R_{n_{1}+1}^{\prime }\left( Y,x_{n}\right) =-T_{n_{2}+1}\left(
Q,x_{n}\right) T_{n_{1}-n_{2}}\left( P\right).  \label{eq 0.7}
\end{equation}

If $\left\{ Q,x_{n}\right\} <P$, $\left\vert Q\right\vert =n_{2}$%
, we have that%
\begin{equation}
A_{n_{1}+1}^{\prime }\left( Y,x_{n}\right) =-T_{n_{1}-n_{2}}\left( P\right)
T_{n_{2}+1}\left( Q,x_{n}\right) .
\end{equation}
\end{theorem}
The validity of this theorem is guaranteed for distributions with more than two points. \footnote{The proof of this theorem follows similar steps given in the Theorem 1.1 in Ref.\cite{ref19}, but now $ > $ indicates the causal relation in the light-front form.}

Moreover, another important distributions are obtained if the sums of Eq.\eqref{eq 0.12}, \eqref{eq 0.13} are extended over all partitions $P_{2}^{0}$, including the empty set $X=\emptyset $, these are the \emph{advanced} and \emph{retarded} distributions
\begin{align}
A_{n}\left( x_{1},...,x_{n}\right)& \equiv  \sum_{P_{2}^{0}}\tilde{T}%
_{n_{1}}\left( X\right) T_{n-n_{1}}\left( Y,x_{n}\right) ,\notag \\
 &=A_{n}^{\prime }\left(x_{1},...,x_{n}\right) +T_{n}\left( x_{1},...,x_{n}\right),  \label{eq 0.9} \\
R_{n}\left( x_{1},...,x_{n}\right) &\equiv \sum_{P_{2}^{0}}T_{n-n_{1}}\left(
Y,x_{n}\right) \tilde{T}_{n_{1}}\left( X\right),\notag \\
 &= R_{n}^{\prime }\left(x_{1},...,x_{n}\right) +T_{n}\left( x_{1},...,x_{n}\right) . \label{eq 0.11}
\end{align}
We see that these two distributions have an extra term when compared with $A_{n}^{\prime }$ and $R_{n}^{\prime }$, respectively, and it is precisely because of this term that these two distributions are not known by the induction assumption.

It should be emphasized that either $R_n$ or $A_n$ can be determined separately by investigating the support properties of various distributions, this is precisely the crucial point where the causal structure becomes very important. Moreover, one may conclude from the theorem 2.1 and causal properties, Eq.\eqref{eq 1.6}, that $R_{n}$ is a retarded and $A_{n}$ an advanced distribution
\begin{equation}
\text{Supp}~R_{n}\left( x_{1},...,x_{n}\right) \subseteq \Gamma _{n-1}^{+}\left(x_{n}\right) , \qquad \text{Supp}~A_{n}\left( x_{1},...,x_{n}\right) \subseteq \Gamma _{n-1}^{-}\left(x_{n}\right) ,
\end{equation}%
where
\begin{align}
\Gamma _{n-1}^{\pm}\left( x_{n}\right) & \equiv \left\{ \left(x_{1},...,x_{n}\right) | ~ x_{j}\in \bar{V}^{\pm}\left(
x_{n}\right) ,~ \forall ~ j=1,...,n-1\right\} , \\
\bar{V}^{\pm}\left( x_{n}\right) &=\left\{ y ~ | \left( y^{+}-x_{n}^{+}\right) \geq 0, ~ \pm \left( y^{-}-x_{n}^{-}\right) \geq 0\right\},
\end{align}
and $\bar{V}^{\pm }\left(x_{n}\right) $ is the closed forward (backward) cone in the light-front coordinates.

Although the distributions $A_n$ and $R_n$ are not initially known, one may constructed a distribution by the set $\left\{
T_{1},\ldots ,T_{n-1},\tilde{T}_{1},\ldots ,\tilde{T}_{n-1}\right\} $, and it is the so-called \emph{causal distribution} defined as it follows%
\begin{equation}
D_{n}\left( x_{1},...,x_{n}\right) \equiv R_{n}^{\prime }\left(
x_{1},...,x_{n}\right) -A_{n}^{\prime }\left( x_{1},...,x_{n}\right)=R_{n}\left( x_{1},...,x_{n}\right)
-A_{n}\left( x_{1},...,x_{n}\right) ,  \label{D def}
\end{equation}
where we have used the relations \eqref{eq 0.9} and \eqref{eq 0.11} between $A_{n}^{\prime }$, $A_{n}$, and $R_{n}^{\prime }$, $R_{n}$, respectively. Furthermore, it follows that, from the theorem 2.1 and causal properties, we can conclude that the support of $D_{n}$ has \emph{causal support} respect to $x_{n}$
\begin{equation}
\text{Supp}~D_{n}\left( x_{1},...,x_{n}\right) \subseteq \Gamma _{n-1}^{+}\left(
x_{n}\right) \cup \Gamma _{n-1}^{-}\left( x_{n}\right) .
\end{equation}
Finally we stress that the above mentioned distributions Eq.\eqref{D def} can be constructed from the set $\left\{ T_{1},\ldots ,T_{n-1},\tilde{T}_{1},\ldots ,\tilde{T}_{n-1}\right\} $ and from that we can make contact with the desired $T_{n}$ distribution via%
\begin{equation}
T_{n}=A_{n}-A_{n}^{\prime }=R_{n}-R_{n}^{\prime }.
\end{equation}
It should be emphasized, however, that all products of distributions in here are mathematically well-defined quantities
because of its arguments are disjoint sets of points in such a way that the their product are tensor products of distributions.

\subsection{Distribution splitting}

For Light-Front Quantum Electrodynamics, like the usual QED, the general form of the causal distribution \eqref{D def} can be written as the following normally ordered product
\begin{equation}
D_{n}\left( x_{1},...,x_{n}\right) =\sum\limits_{k}d_{n}^{k}\left(
x_{1},...,x_{n}\right) :\prod\limits_{j}\bar{\psi}\left( x_{j}\right)
\prod\limits_{l}\psi \left( x_{l}\right) \prod\limits_{m}A\left(
x_{m}\right) :
\end{equation}%
where $d_{n}^{k}\left( x_{1},...,x_{n}\right) $ is its numerical part. \footnote{Because of translational invariance, we have
that $d_{n}^{k}$ depends only on relative coordinates: $ d\left( x\right) \equiv d_{n}^{k}\left( x_{1}-x_{n},...,x_{n-1}-x_{n}\right).$}

The crucial point of the inductive process is the splitting problem of the
distribution $d$ at the origin $ \left\{ x_{n}\right\} $ into a (retarded) distribution $r$ with support in $ \Gamma _{n-1}^{+}\left( x_{n}\right) $ and a (advanced) distribution $a$ with support in $\Gamma_{n-1}^{-}\left( x_{n}\right) $.\footnote{In fact, this problem of distributional splitting is a well-known feature established by Malgrange in a general framework \cite{ref47}.} For this purpose we need to classify the distribution $d$ at the origin $ x=0 $ or, in the momentum space, at $ p=\infty $ \cite{ref19}.
Thus, it is said that the distribution $d\left( x\right) \in \mathcal{J}^{\prime }\left(\mathbb{R}^{m}\right) $ has the singular order $ \omega $ if its Fourier transform $\hat{d}\left( p\right) $ $\in \mathcal{J}^{\prime }\left(\mathbb{R} ^{m}\right) $ has a quasi-asymptotics $\hat{d}_{0}\left( p\right) $ at $p=\infty $ with regard to a positive continuous function $\rho \left( \alpha
\right) $, $\alpha >0$, if the limit%
\begin{equation}
\lim_{\alpha \rightarrow 0}\rho \left( \alpha \right) \hat{d}\left( \frac{p}{%
\alpha }\right) =\hat{d}_{0}\left( p\right) \neq 0,
\end{equation}%
exists in $\mathcal{J}^{\prime }\left(\mathbb{R} ^{m}\right) $, here $ m=4(n-1) $ is the dimension of the Schwartz space; whereas the \emph{power-counting function} $\rho \left( \alpha\right) $ satisfying
\begin{equation}
\lim_{\alpha \rightarrow 0}\frac{\rho \left( a\alpha \right) }{\rho \left(
\alpha \right) }=a^{\omega }, \quad \forall ~ a>0. \label{sing}
\end{equation}

By requiring that the splitting procedure preserves the singular order of the distributions we have two distinguished cases:

(i) \emph{Regular distributions}, for this case $\omega<0 $, and retarded distribution $ r(x) $ can be obtained by multiplication of the causal distribution $ d(x) $ by a $\theta $-Heaviside function as follows
\begin{equation}
r\left( x\right) =\theta \left( v.x\right) d\left( x\right) , \label{eq 0.17a}
\end{equation}%
where $v=\left( v_{1},\ldots ,v_{n-1}\right) \in \Gamma ^{+}$ , which guarantees $v.x\geq 0$ for all $x$ inside the forward light-cone $\Gamma ^{+}$. By means of convenience in the calculation, the product \eqref{eq 0.17a} can be rewritten in the momentum space as the following \footnote{One should remember that the product in $x$-space of a tempered distribution and a test function in Schwartz space goes over into a convolution in $p$-space.}
\begin{equation}
\hat{r}\left( p\right) =\left( 2\pi \right) ^{-2}\int dk\hat{\theta}\left(
p-k\right) \hat{d}\left( k\right) .
\end{equation}
In Light-Front dynamics we may choose, in particular: $v=\left( 1,0^{\bot },1;0;\ldots ;0\right) $ then it implies into $\theta \left( v.x\right) =\theta \left( x_{1}^{+}+x_{1}^{-}\right) $, or in the momentum space:
\begin{equation}
\hat{\theta}\left( k\right) =\left( 2\pi \right) ^{\frac{m}{2}-1}\delta
\left( k_{1}^{\bot },k_{2},\ldots ,k_{m}\right) \frac{i}{\kappa ^{+}+i0^{+}}%
\delta \left( \kappa _{1}^{-}-\kappa _{1}^{+}\right) .
\end{equation}
Now, in addition to the above result we may also choose a coordinate system such that $ p=\left( p^{\prime },0^{\bot },p^{\prime };0;\ldots ;0\right) $, i.e., taking $p$ to be parallel to $v$. This leads to the dispersion relation form for the retarded distribution
\begin{equation}
\hat{r}\left( p^{\prime },0^{\bot },p^{\prime }\right) =\frac{i}{2\pi }\int
dk^{-}dk^{+}\frac{\hat{d}\left( k^{-},0^{\bot },k^{+}\right) }{p^{\prime }-k^{+}+i0^{+}}\delta \left( k^{+}-k^{-}\right).
\end{equation}
Moreover, prescribing the support of the distribution $\delta $-Dirac in such a way that: $k^{+}=k^{-}=k $, and also defining the variable of integration $t=k/p^{\prime }$, we find that
\begin{equation}
\hat{r}\left( p^{\prime },0^{\bot },p^{\prime }\right) =\frac{i}{2\pi }%
sgn\left( p^{\prime }\right) \int\limits_{-\infty }^{\infty }dt\frac{\hat{d}%
\left( tp^{\prime },0^{\bot },tp^{\prime }\right) }{1-t+sgn\left( p^{\prime
}\right) i0^{+}}.
\end{equation}
Finally, to write down this result for an arbitrary four-vector $p\in \Gamma ^{+}\cup \Gamma ^{-}$ we must apply a boost and rotation transformation, the resulting expression then reads
\begin{equation}
\hat{r}\left( p\right) =\frac{i}{2\pi }sgn\left( p_{\lambda }\right)
\int\limits_{-\infty }^{\infty }dt\frac{\hat{d}\left( tp\right) }{1-t+sgn\left( p_{\lambda }\right) i0^{+}} \label{eq 1.41},
\end{equation}%
this is a \emph{dispersion relation without subtractions}. Also, $p_{\lambda }$ is a parameter of some time-like curve, or inclusive light-like curves, which passes by the origin. In particular we can choose $p_{\lambda }=p^{+}$ or equivalently $p_{\lambda }=p^{-}$. \footnote{But we must avoid ill-defined products like $sgn\left( p^{-}\right) \delta \left(
p^{-}\right) $.}

(ii) \emph{Singular distributions}, for this case $ \omega\geqslant0 $, and the retarded distribution $ r(x) $ is defined as
\begin{equation}
\left\langle r\left( x\right) ,\varphi \left( x\right) \right\rangle=
\left\langle \theta \left( v.x\right) d\left( x\right) ,\mathcal{W}\varphi
\left( x\right) \right\rangle ,
\end{equation}%
where $\mathcal{W}$ is an projector operator over the original test function space ${\varphi} $ \cite{ref19}.  In fact, we see that the careless multiplication $\theta \left( v.x\right) d\left( x\right) $ in $p$-space is ill-defined in this singular case, and it yields to an ultraviolet divergent expression. Nevertheless, following similar steps as those from the regular case, and considering the following normalization condition at the origin: $D^{b}\hat{r} \left( p \right) =0, ~\forall ~ \left\vert b\right\vert \leq \omega $ we can find that, for an arbitrary $p\in \Gamma ^{+}\cup \Gamma ^{-}$, the retarded distribution is expressed as:%
\begin{equation}
\hat{r}\left( p\right) =\frac{i}{2\pi }sgn\left( p_{\lambda }\right)
\int\limits_{-\infty }^{+\infty }dt \frac{\hat{d}\left( tp\right) }{%
t^{\omega +1}\left( 1-t+sgn\left( p_{\lambda }\right) i0^{+}\right) }, \label{eq 1.40}
\end{equation}%
this is a \emph{dispersion relation} with $\omega +1$ \emph{subtractions}. From the normalization condition we call Eq.\eqref{eq 1.40} the \emph{central splitting solution}; besides, it is known that the central splitting solution preserves most of the original symmetries of the theory, such as Lorentz covariance.

In order to conclude our development, one may now define a new valid retarded distribution solution of the form \cite{ref27}
\begin{equation}
\tilde{r}\left( p\right) =\hat{r}\left( p\right) +\sum_{\left\vert
a\right\vert =0}^{\omega }C_{a}p^{a}.
\end{equation}
So we are left with free coefficients $C_{a}$ that can not be determined by the causal structure, but they must be restricted by further physical considerations.\footnote{In the causal approach, this procedure is known as \textit{polynomial normalization} and it is somehow related to the usual renormalization procedure of the usual perturbative QFT.}

\section{Vacuum Polarization}

\label{sec:2}

The perturbative program has its start when we first construct the intermediate distributions:
\begin{equation}
R_{2}^{\prime }\left( x_{1},x_{2}\right) =-T_{1}\left( x_{2}\right)
T_{1}\left( x_{1}\right) ,\quad A_{2}^{\prime }\left( x_{1},x_{2}\right)
=-T_{1}\left( x_{1}\right) T_{1}\left( x_{2}\right) ,  \label{eq 2.1}
\end{equation}%
and subsequently the causal distribution $D_{2}$ as it follows%
\begin{equation}
D_{2}\left( x_{1},x_{2}\right) =R_{2}^{\prime }\left( x_{1},x_{2}\right)
-A_{2}^{\prime }\left( x_{1},x_{2}\right) =\left[ T_{1}\left( x_{1}\right)
,T_{1}\left( x_{2}\right) \right] .  \label{eq 2.2}
\end{equation}%
For QED$_{4}$ we consider as the first perturbative term: $T_{1}\left(x\right) =ie\colon \bar{\psi}\left( x\right) \gamma ^{\mu }\psi \left(x\right) \colon A_{\mu }\left( x\right) $. Thus, after applying the Wick theorem for normally ordering products, we obtain from all of these terms those associated with the vacuum polarization (VP) contributions:%
\begin{align}
R_{2}^{\prime VP}\left( x_{1},x_{2}\right) &= e^{2}\colon A_{\nu }\left(
x_{2}\right) \overbrace{\bar{\psi}\left( x_{2}\right) \gamma ^{\nu }%
\overbrace{\psi \left( x_{2}\right) \bar{\psi}\left( x_{1}\right) }\gamma
^{\mu }\psi \left( x_{1}\right) }A_{\mu }\left( x_{1}\right) \colon ,
\label{eq 2.3} \\
A_{2}^{\prime VP}\left( x_{1},x_{2}\right) &= e^{2}\colon A_{\mu }\left(
x_{1}\right) \overbrace{\bar{\psi}\left( x_{1}\right) \gamma ^{\mu }%
\overbrace{\psi \left( x_{1}\right) \bar{\psi}\left( x_{2}\right) }\gamma
^{\nu }\psi \left( x_{2}\right) }A_{\nu }\left( x_{2}\right) \colon .
\label{eq 2.4}
\end{align}%
Moreover, we have that the fermionic contractions are defined as follows \cite{ref28}
\begin{equation}
\overbrace{\psi _{a}\left( x_{1}\right) \bar{\psi}_{b}\left( x_{2}\right) }
=-iS_{ab}^{\left( +\right) }\left( x_{1}-x_{2}\right) , \quad
\overbrace{\bar{\psi}_{a}\left( x_{1}\right) \psi _{b}\left( x_{2}\right) }
=-iS_{ba}^{\left( -\right) }\left( x_{2}-x_{1}\right) ,
\end{equation}%
where $S^{\left( +\right) }$ and $S^{\left( -\right) }$ are the positive (PF) and negative (NF) frequency parts of the fermionic propagator, respectively. After some calculation, and as well as by introducing the tensor $P^{\mu \nu }\left( y\right) =e^{2}tr\left[ \gamma ^{\mu }S^{\left(+\right) }\left( y\right) \gamma ^{\nu }S^{\left( -\right) }\left( -y\right)
\right] $, we arrive at the expression:
\begin{align}
R_{2}^{\prime VP}\left( x_{1},x_{2}\right) &= -\colon A_{\mu }\left(
x_{1}\right) P^{\nu \mu }\left( x_{2}-x_{1}\right) A_{\nu }\left(
x_{2}\right) \colon  \label{eq 2.5} \\
D_{2}^{VP}\left( x_{1},x_{2}\right) &= \colon A_{\mu }\left( x_{1}\right)
d^{\mu \nu }\left( x_{1}-x_{2}\right) A_{\nu }\left( x_{2}\right) \colon ,
\label{eq 2.6}
\end{align}%
with $d^{\mu \nu }\left( y\right) \equiv P^{\mu \nu }\left( y\right) -P^{\nu\mu }\left( -y\right) $.

As we have mentioned earlier it is necessary to prove that $D_{2}^{VP}$, rather its numerical part $d^{\mu \nu }$, has causal support. Hence, it follows that, after some manipulation, $d^{\mu \nu }$ can be written as%
\begin{equation}
e^{-2}d^{\mu \nu }\left( x_{1}-x_{2}\right) =tr\left\{ \gamma ^{\mu}S^{\left( +\right) }\left( x_{1}-x_{2}\right) \gamma ^{\nu }S\left(x_{2}-x_{1}\right) \right\} -tr\left\{ \gamma ^{\mu }S\left(x_{1}-x_{2}\right) \gamma ^{\nu }S^{\left( +\right) }\left(x_{2}-x_{1}\right) \right\} .  \label{eq 2.7}
\end{equation}%
We have that, in the light-front, the fermionic causal propagator $S$ has the form \cite{ref28}
\begin{equation}
S\left( x\right) =\left( i\gamma .\partial +m\right) D_{m}\left( x\right) ,\label{eq 2.8}
\end{equation}%
where $D_{m}$ is the scalar causal propagator (Pauli-Jordan distribution):
\begin{equation}
D_{m}\left( x\right) =\frac{sgn\left( x^{-}\right) }{2\pi }\left[ \delta \left( x^{2}\right) -\frac{m}{2}\frac{\theta \left( x^{2}\right) }{\sqrt{x^{2}}}J_{1}\left( m\sqrt{x^{2}}\right) \right] ,  \label{eq 2.9}
\end{equation}%
therefore, since $S$ has causal support, it follows that the product $\left( S^{\left( +\right) }S\right) $ has causal support as well. In this way, we may conclude that the distribution $d^{\mu \nu }$ has causal support, as required; this means: $\text{Supp}~D_{2}^{VP}\left( x_{1},x_{2}\right) \subseteq \Gamma _{2}^{+}\left( x_{2}\right) \cup \Gamma _{2}^{-}\left(x_{2}\right) $. So far, we have not seen any major difference of our results with those from
the usual coordinates \cite{ref19}. However, we should proceed to confirm if this is true until the end, we review each one of the steps of the causal approach.

\subsection{Singular order}

We shall now calculate the singular order following the criterion \eqref{sing} in the momentum space. Thus, first, we need to know the expression $d^{\mu \nu }$ in the momentum space: $\hat{d}^{\mu \nu }$. From the Fourier transform:%
\begin{equation}
d^{\mu \nu }\left( y\right) =\left( 2\pi \right) ^{-2}\int dk\left[ \mathcal{%
F}\left[ P^{\mu \nu }\right] \left( k\right) -\mathcal{F}\left[
P^{\nu \mu }\right] \left( -k\right) \right] e^{-iky},  \label{eq 2.10}
\end{equation}%
it follows that the $\hat{d}^{\mu \nu }$ has the form%
\begin{equation}
\hat{d}^{\mu \nu }\left( k\right) =\hat{P}^{\mu \nu }\left( k\right) -\hat{P}%
^{\nu \mu }\left( -k\right) ,  \label{eq 2.11}
\end{equation}%
where $\hat{P}^{\mu \nu }\left( k\right) \equiv \mathcal{F}\left[ P^{\mu \nu }\right] \left( k\right) $. Then, in order to determine $\hat{d}^{\mu \nu }$ we need to calculate the Fourier transformation of $P^{\mu \nu }$. From the explicit definition of $P^{\mu \nu }$, we have that its Fourier transformation is:
\begin{equation}
\hat{P}^{\mu \nu }\left( k\right) =\mathcal{F}\left[ P^{\mu \nu }\right] \left( k\right) =e^{2}\left( 2\pi \right) ^{-2}\int dytr\left[ \gamma ^{\mu }S^{\left( +\right) }\left( y\right) \gamma ^{\nu }S^{\left( -\right)
}\left( -y\right) \right] e^{iky}.  \label{eq 2.12}
\end{equation}%
Moreover, replacing the Fourier expansion for the fermionic PF and NF propagators \cite{ref28}, and after some manipulation, we obtain that
\begin{equation}
\hat{P}^{\mu \nu }\left( k\right) =-e^{2}\left( 2\pi \right) ^{-2}\int d^{4}ptr\left[ \gamma ^{\mu }\left( \gamma .p+m\right) \gamma ^{\nu }\left( \gamma .k-\gamma .p-m\right) \right] \hat{D}_{m}^{\left( +\right) }\left(
p\right) \hat{D}_{m}^{\left( -\right) }\left( p-k\right) .  \label{eq 2.14}
\end{equation}%
From the general trace properties of the $\gamma $-matrices, we obtain:
\begin{equation}
tr\left[ \gamma ^{\mu }\left( \gamma .p+m\right) \gamma ^{\nu }\left( \gamma .k-\gamma .p-m\right) \right] =4\left[ p^{\mu }k^{\nu }+p^{\nu }k^{\mu }-2p^{\mu }p^{\nu }-h^{\mu \nu }\left( \left( p.k\right) -\left(p^{2}-m^{2}\right) \right) \right] .  \label{eq 2.15}
\end{equation}%
Finally, one may use the distributional property of the $\delta $-Dirac, $x\delta\left(x\right)=0$, to show that $\hat{P}^{\mu \nu }$ can be expressed as%
\begin{equation}
\hat{P}^{\mu \nu }\left( k\right) =-4e^{2}\left( 2\pi \right) ^{-2}\int d^{4}p\left[ p^{\mu }k^{\nu }+p^{\nu }k^{\mu }-2p^{\mu }p^{\nu }-h^{\mu \nu }\left( p.k\right) \right] \hat{D}_{m}^{\left( +\right) }\left( p\right)
\hat{D}_{m}^{\left( -\right) }\left( p-k\right) .  \label{eq 2.16}
\end{equation}%
After some simple calculation, we can prove that the tensor $\hat{P}^{\mu \nu }$ \eqref{eq 2.16} satisfies
\begin{equation}
k_{\mu }\hat{P}^{\mu \nu }\left( k\right) =0,  \label{eq 2.17}
\end{equation}%
which means that the vacuum polarization \eqref{eq 2.6} is gauge-invariant. Moreover, this result shows that $\hat{P}^{\mu \nu }\left( k\right) $ is a transversal tensor, which can be cast in the following form
\begin{equation}
\hat{P}^{\mu \nu }\left( k\right) =e^{2}\left( 2\pi \right) ^{-4}\left( h^{\mu \nu }-\frac{k^{\mu }k^{\nu }}{%
k^{2}}\right) \hat{d}_{1}\left( k\right) .  \label{eq 2.18}
\end{equation}%
Using the spherical symmetry, the scalar distribution $\hat{d}_{1}\left( k\right) $ reads
\begin{equation}
\hat{d}_{1}\left( k\right) =\frac{4}{3}\left( 2m^{2}+k^{2}\right) \left(2\pi \right) ^{2}\int d^{4}p\hat{D}_{m}^{\left( +\right) }\left( p\right) \hat{D}_{m}^{\left( -\right) }\left( p-k\right) .  \label{eq 2.19}
\end{equation}%
In this expression the integral is proportional to the convolution: $\left[ \hat{D}_{m}^{\left( +\right) }\ast \hat{D}_{m}^{\left( -\right) }\right] \left( k\right) $, \textit{a priori} we do not know if the result is the
same that the usual coordinates. The evaluation of this integral \footnote{This convolution is explicitly calculated in the \ref{app:B}.} is straightforward and we obtain that:%
\begin{equation}
\hat{d}_{1}\left( k\right) =\frac{2\pi }{3}\left( 2m^{2}+k^{2}\right) \theta
\left( k^{-}\right) \theta \left[ k^{2}-4m^{2}\right] \sqrt{1-\frac{4m^{2}}{%
k^{2}}}.  \label{eq 2.20}
\end{equation}%
We may now substitute these results into the Eq.\eqref{eq 2.10}, and then obtain the numerical causal distribution $d^{\mu \nu }$ in the momentum space:%
\begin{equation}
\hat{d}^{\mu \nu }\left( k\right) =e^{2}\left( 2\pi \right) ^{-4}\left(
h^{\mu \nu }-\frac{k^{\mu }k^{\nu }}{k^{2}}\right) \hat{d}\left( k\right) ,\label{eq 2.21}
\end{equation}%
where
\begin{equation}
\hat{d}\left( k\right) \equiv \hat{d}_{1}\left( k\right) -\hat{d}_{1}\left( -k\right) =\frac{2\pi }{3}\left(2m^{2}+k^{2}\right) sgn\left( k^{-}\right)\theta \left( k^{2}-4m^{2}\right) \sqrt{1-\frac{4m^{2}}{k^{2}}}.
\label{eq 2.22}
\end{equation}%
Before starting the splitting procedure, we must first determine the singular order $\omega $ of this distribution, this can be obtained from the expression $\hat{d}_{\mu \nu }\left( \frac{k}{\alpha }\right) $ when $\alpha\rightarrow 0^{+}$, using the previous result:%
\begin{align}
\hat{d}^{\mu \nu }\left( \frac{k}{\alpha }\right) =&\alpha ^{-2}e^{2}\left(
2\pi \right) ^{-4}\left( h^{\mu \nu }-\frac{k^{\mu }k^{\nu }}{k^{2}}\right)\notag \\
& \quad \times \left[ \frac{2\pi }{3}\left( 2m^{2}\alpha ^{2}+k^{2}\right) \theta \left( k^{2}-4m^{2}\alpha ^{2}\right) sgn\left( \frac{k^{-}}{\alpha }\right)\left( \sqrt{1-\frac{4m^{2}\alpha ^{2}}{k^{2}}}\right) \right] , \notag \\
& \rightarrow  \alpha ^{-2}e^{2}\left( 2\pi \right) ^{-4}\left( h^{\mu \nu }-\frac{k^{\mu }k^{\nu }}{k^{2}}\right) \left[ \frac{2\pi }{3}\left(k^{2}\right) \theta \left( k^{2}\right) sgn\left( k^{-}\right) \right] .\label{eq 2.23}
\end{align}%
Hence, from the above result we may say that the vacuum polarization at one-loop has singular order:%
\begin{equation}
\omega _{2}^{VP}=+2.
\end{equation}%
This result is usually related to the power counting degree of the usual instant form QED$_4$, which has the same value for this case: $+2$. But as it has been showed by A. Aste \textit{et al} \cite{ref20} in the case of the Schwinger model, in general, this is not always true. For the causal approach determining carefully the singular order is mandatory.

\subsection{Retarded part of the vacuum polarization at one-loop}

Since $\hat{d}^{\mu \nu }$ is a distribution of singular order $+2$, we should use the following splitting formula \eqref{eq 1.40}, in order to obtain the retarded distribution $\hat{r}^{\mu \nu }$:%
\begin{equation}
\hat{r}^{\mu \nu }\left( k\right) =\frac{i}{2\pi }sgn\left( k^{-}\right)
\int\limits_{-\infty }^{+\infty }dt\frac{\hat{d}^{\mu \nu }\left( tk\right)
}{t^{3}\left( 1-t+sgn\left( k^{-}\right) i0^{+}\right) }.  \label{eq 2.24}
\end{equation}%
Furthermore, substituting the expression \eqref{eq 2.21} of the causal distribution into the formula \eqref{eq 2.24}, we obtain that%
\begin{equation}
\hat{r}^{\mu \nu }\left( k\right) =e^{2}\left( 2\pi \right) ^{-4}\left(
h^{\mu \nu }-\frac{k^{\mu }k^{\nu }}{k^{2}}\right) \left[ \frac{i}{2\pi }%
sgn\left( k^{-}\right) \int\limits_{-\infty }^{+\infty }dt\frac{\hat{d}%
\left( tk\right) }{t^{2+1}\left( 1-t+sgn\left( k^{-}\right) i0^{+}\right) }%
\right] .  \label{eq 2.25}
\end{equation}%
Because the tensor character of \eqref{eq 2.25} we may focus our attention in solving
\begin{equation}
\hat{r}\left( k\right) =\frac{i}{2\pi }sgn\left( k^{-}\right)
\int\limits_{-\infty }^{+\infty }dt\frac{\hat{d}\left( tk\right) }{%
t^{3}\left( 1-t+sgn\left( k^{-}\right) i0^{+}\right) }.  \label{eq 2.26}
\end{equation}%
Hence, substituting the expression \eqref{eq 2.22} for $\hat{d}$ and, since this is an odd function in $t$, we obtain that%
\begin{equation}
\hat{r}\left( k\right) =\frac{i}{3}\int\limits_{\frac{4m^{2}}{k^{2}}%
}^{\infty }ds\frac{1}{1-s+sgn\left( k^{-}\right) i0^{+}}\left( \frac{%
2m^{2}+sk^{2}}{s^{2}}\right) \left( \sqrt{1-\frac{4m^{2}}{sk^{2}}}\right) ,
\label{eq 2.27}
\end{equation}%
where we had made the substitution $t^{2}\rightarrow s$. Moreover, after some manipulation, \footnote{By means of the Sochozki formula: $\frac{1}{x+sgn\left( k^{-}\right) i0^{+}}=\frac{1}{x+i0^{+}}+2i\pi \theta \left( -k^{-}\right) \delta \left( x\right) $, see Ref.\cite{ref27}.} then $\hat{r}$ can also be written as%
\begin{align}
\hat{r}\left( k\right) &=\frac{i}{3}\int\limits_{\frac{4m^{2}}{k^{2}}%
}^{\infty }ds\frac{1}{1-s+i0^{+}}\left( \frac{2m^{2}+sk^{2}}{s^{2}}\right)
\sqrt{1-\frac{4m^{2}}{sk^{2}}}  \notag \\
& \quad-\frac{2}{3}\theta \left( -k^{-}\right) \theta \left[ k^{2}-4m^{2}\right]
\left( 2m^{2}+k^{2}\right) \sqrt{1-\frac{4m^{2}}{k^{2}}}.  \label{eq 2.29}
\end{align}%
Recalling the result \cite{ref19}%
\begin{align}
I\left( k\right) &\equiv \int\limits_{\frac{4m^{2}}{k^{2}}}^{\infty }ds \frac{1}{1-s+i0^{+}}\left( \frac{2m^{2}+sk^{2}}{s^{2}}\right) \sqrt{1-\frac{4m^{2}}{sk^{2}}} , \notag \\
&=m^{2}\left[ \frac{1+\xi }{1-\xi }\left( \xi -4+\frac{1}{\xi }\right) \ln
\xi +\frac{5}{3\xi }+\frac{5\xi }{3}-\frac{22}{3}\right] .  \label{eq 2.30}
\end{align}%
where $\xi ^{\pm 1}=\left( 1-\frac{k^{2}}{2m^{2}}\right) \pm \frac{k^{2}}{2m^{2}}\sqrt{1-\frac{4m^{2}}{k^{2}}}$. Then, we arrive at the following explicit expression for the retarded part of the vacuum polarization \footnote{The only difference in respect to the instant-form expression \cite{ref19} is in $\hat{d}_{1}\left( -k\right) $, but this corresponds in replacing $\theta \left( k^{-}\right)\rightarrow \theta \left( k^{0}\right) $ in its definition.}
\begin{align}
\hat{r}\left( k\right) &=\frac{i}{3}m^{2}\left\{ \frac{1+\xi }{1-\xi }%
\left( \xi -4+\frac{1}{\xi }\right) \ln \xi +\frac{5}{3\xi }+\frac{5\xi }{3}-%
\frac{22}{3}\right\}  \notag \\
& \quad-\frac{2}{3}\theta \left( -k^{-}\right) \theta \left[ k^{2}-4m^{2}\right]
\left( 2m^{2}+k^{2}\right) \sqrt{1-\frac{4m^{2}}{k^{2}}}.  \label{eq 2.31}
\end{align}%
Finally, we conclude that the retarded part of the vacuum polarization tensor \eqref{eq 2.25} has the form%
\begin{equation}
\hat{r}^{\mu \nu }\left( k\right) =e^{2}\left( 2\pi \right) ^{-4}\left(
h^{\mu \nu }-\frac{k^{\mu }k^{\nu }}{k^{2}}\right) \hat{r}\left( k\right) .
\label{eq 2.32}
\end{equation}

\subsection{Vacuum polarization tensor at one-loop}

Although it follows several terms from the expression \eqref{eq 2.2} we can focus our attention only in those terms associated with the vacuum polarization contribution: $T_{2}^{VP}\left( x_{1},x_{2}\right) $. This contribution is
obtained from the relation:%
\begin{equation}
T_{2}^{VP}\left( x_{1},x_{2}\right) =R_{2}^{VP}\left( x_{1},x_{2}\right)
-R_{2}^{\prime VP}\left( x_{1},x_{2}\right) ,  \label{eq 2.33}
\end{equation}%
where $R_{2}^{VP}$ is the retarded part of $D_{2}^{VP}$. From our previous results we have that%
\begin{align}
R_{2}^{\prime VP}\left( x_{1},x_{2}\right) &= \,-\colon A_{\mu }\left(
x_{1}\right) P^{\nu \mu }\left( x_{2}-x_{1}\right) A_{\nu }\left(
x_{2}\right) \colon , \\
R_{2}^{VP}\left( x_{1},x_{2}\right) &= \,\colon A_{\mu }\left( x_{1}\right)
r^{\mu \nu }\left( x_{1}-x_{2}\right) A_{\nu }\left( x_{2}\right) \colon .
\end{align}%
Then the complete contribution $T_{2}^{VP}$ can be written in the form:%
\begin{equation}
T_{2}^{VP}\left( x_{1},x_{2}\right) =-i\colon A_{\mu }\left( x_{1}\right)
\Pi ^{\mu \nu }\left( x_{1}-x_{2}\right) A_{\nu }\left( x_{2}\right) \colon ,
\label{eq 2.34}
\end{equation}%
where $\Pi ^{\mu \nu }$ is the known \emph{vacuum polarization tensor}, and it is defined by the relation%
\begin{equation}
\Pi ^{\mu \nu }\left( x_{1}-x_{2}\right) =i\left[ r^{\mu \nu }\left(
x_{1}-x_{2}\right) +P^{\nu \mu }\left( x_{2}-x_{1}\right) \right] ,\label{eq 2.35}
\end{equation}%
or rather in the momentum space%
\begin{equation}
\hat{\Pi}^{\mu \nu }\left( k\right) =i\left[ \hat{r}^{\mu \nu }\left(k\right) +\hat{P}^{\nu \mu }\left( -k\right) \right] .  \label{eq 2.36}
\end{equation}%
Replacing the expressions of $\hat{P}^{\mu \nu }$ and $\hat{r}^{\mu \nu }$, Eqs.\eqref{eq 2.18} and \eqref{eq 2.32}, respectively, we obtain that $\hat{\Pi}^{\mu \nu }$ may be written as the following%
\begin{equation}
\hat{\Pi}^{\mu \nu }\left( k\right) =-\left( 2\pi \right) ^{-4}\left( h^{\mu
\nu }-\frac{k^{\mu }k^{\nu }}{k^{2}}\right) \ \hat{\Pi}\left( k\right) ,\label{eq 2.37}
\end{equation}%
this clearly shows that $\hat{\Pi}^{\mu \nu }$ is a transversal tensor, moreover, we have defined in this expression: $\hat{\Pi}\left( k\right) =-i\left[ \hat{r}\left( k\right) +\hat{d}_{1}\left( -k\right) \right] $. This quantity is the so-called \emph{vacuum polarization scalar}. Furthermore, replacing $\hat{r}\left( k\right) $ and $\hat{d}_{1}\left( -k\right) $, from Eqs.\eqref{eq 2.31} and \eqref{eq 2.20}, respectively, we obtain that%
\begin{equation}
\hat{\Pi}\left( k\right) =\frac{e^{2}m^{2}}{3}\left\{ \frac{1+\xi }{1-\xi }%
\left( \xi -4+\frac{1}{\xi }\right) \ln \xi +\frac{5}{3\xi }+\frac{5\xi }{3}-%
\frac{22}{3}\right\} .  \label{eq 2.38}
\end{equation}%
Nevertheless, it is known \cite{ref22,ref21,ref24} that this result has different forms and meanings depending on the value of $k^{2}$ such as: the scattering sector for $k^{2}<0$, the unphysical sector for $0<k^{2}<4m^2$ and the production sector for $4m^2 <k^{2}$. Besides, we also see that $\hat{\Pi} \left( k\right) $, Eq.\eqref{eq 2.38}, does not depend on the coordinates system, so the only difference between our light-front vacuum polarization tensor and its instant-form counterpart is the explicit transverse projector in its definition.

As we have mentioned at the end of the Section \ref{sec:1}, we have that for non-negative values of the singular order $\omega$ the solution for this finite perturbation theory is not unique. Since $\omega _{2}^{VP}=+2$, then the following expression is also a solution for the vacuum polarization scalar the following expression%
\begin{equation}
\tilde{\Pi}\left( k\right) =\hat{\Pi}\left( k\right) +C_{0}+C_{\mu }k^{\mu }+C_{1}k^{2},  \label{eq 2.39}
\end{equation}%
in which $C_{0}$, $C_{\mu }$ and $C_{1}$ are constants. To fix these constants we need to consider other physical conditions, additional to those axioms considered initially; for instance, discrete symmetries. Thus, considering
parity, we see that the constant $C_{\mu }$ must vanish,
\begin{equation}
\tilde{\Pi}\left( k\right) =\hat{\Pi}\left( k\right) +C_{0}+C_{1}k^{2}.
\label{eq 2.40}
\end{equation}
The remaining constants $C_0$ and $C_1$ are obtained when we analyze the complete photon propagator expression, modified
by vacuum polarization insertions, in the one-loop approximation. This is given by the series
\begin{equation}
\left[\hat{\mathcal{D}}_{\mu \nu }^{F}\left( k\right)\right]^{-1}=\left[\hat{D}_{\mu \nu }^{F}\left( k\right)\right]^{-1}
-\left(2\pi\right)^{4}\tilde{\Pi}_{\mu \nu }\left( k\right),  \label{eq 2.41a}
\end{equation}%
where $\hat{D}_{\mu \nu }^{F}\left( k\right)$ is the free photon propagator \cite{ref28}. Thus we obtain that the complete double transverse \footnote{By doubly transverse we mean that $k^{\mu}\hat{D}_{\mu \nu }^{F}\left( k\right)=0$ and
$\eta^{\mu}\hat{D}_{\mu \nu }^{F}\left( k\right)=0$.} photon propagator takes the form
\begin{equation}
\hat{\mathcal{D}}_{\mu \nu }^{F}\left( k\right) =\left\{
\frac{h_{\mu \nu }}{k^{2}+i0^{+}}-\frac{k_{\mu }\eta _{\nu }+k_{\nu }\eta
_{\mu }}{\left( k^{2}+i0^{+}\right) \left[ k^{+}+sgn\left( k^{-}\right)
i0^{+}\right] }+\frac{\eta _{\mu }\eta _{\nu }}{\left[ k^{+}+sgn\left(
k^{-}\right) i0^{+}\right] ^{2}}\right\}\left[1-\left( 2\pi \right) ^{-2}\frac{\tilde{\Pi}\left( k\right)}{k^{2}}\right]^{-1},  \label{eq 2.41}
\end{equation}%
where $\eta ^{\mu }=\left( 0,0,0,1\right) $. Now, it is worth to see that the vacuum polarization behaves at low-energy as $\hat{\Pi}\left( k\right) \approx O\left[ \left( \frac{k^{2}}{m^{2}}\right) ^{2}\right] $, and in order to ensure that the pole residue from the photon propagator holds at $k^{2}=0$ when radiative corrections are considered, we can conclude from the Eqs.\eqref{eq 2.40} and \eqref{eq 2.41} that $C_{0}=0$. Besides, in the causal method the coupling constant
$e$ is the physical charge, so it also follows that: $C_{1}=0$. Therefore, we have fixed all constants and,
find that the original central solution $\hat{\Pi}\left( k\right) $ fulfils every required physical conditions.

\section{Vacuum polarization with Instantaneous fermionic part}

\label{sec:3}

In the perturbative Epstein-Glaser program the basic quantities are the PF and NF propagators of the free fields. In the causal approach, these propagators can be obtained directly from the free field equation. For instance, for the fermionic
field we consider the Dirac equations
\begin{equation}
\mathcal{D}\left( \partial \right) \psi =\left( i\gamma .\partial -m\right)
\psi =0,\quad \bar{\psi}\overleftarrow{\mathcal{D}}\left( \partial \right) =%
\bar{\psi}\left( i\gamma .\overleftarrow{\partial }+m\right) =0.
\label{eq 3.1}
\end{equation}%
Then we obtain that its Green's function is expressed as
\begin{equation}
S\left( p\right) =\frac{\gamma .p+m}{p^{2}-m^{2}}.  \label{eq 3.2}
\end{equation}%
Moreover, using the causal program \cite{ref28} we can find the fermionic PF and NF propagators%
\begin{equation}
\hat{S}^{\left( \pm \right) }\left( p\right) =\left( \gamma .p+m\right) \hat{%
D}_{m}^{\left( \pm \right) }\left( p\right) ,
\end{equation}%
where $\hat{D}_{m}^{\left( \pm \right) }$ are the scalar PF and NF propagators, given by
\begin{equation}
\hat{D}_{m}^{\left( \pm \right) }\left( p\right) =\pm \frac{i}{2\pi }\theta
\left( \pm p^{-}\right) \delta \left( 2p^{+}p^{-}-\omega _{m}^{2}\right)
,\quad \omega _{m}^{2}=p_{\bot }^{2}+m^{2}.  \label{eq 3.3}
\end{equation}%
The same causal program give us the other propagators, thus, by means the covariant splitting formula \eqref{eq 1.41} we can obtain the fermionic Feynman propagator%
\begin{equation}
\hat{S}^{F}\left( p\right) =-\left( 2\pi \right) ^{-2}\frac{\gamma .p+m}{%
p^{2}-m^{2}+i0^{+}},\quad p^{2}>0.  \label{eq 3.4}
\end{equation}%
As it may be seen, in order to derive this result we only had to take into account the Dirac equation. However, in the light-front literature, the fermionic Feynman propagator is given by \cite{ref12}
\begin{equation}
\hat{S}^{F}\left( p\right) =-\left( 2\pi \right) ^{-2}\frac{\gamma .p+m}{%
p^{2}-m^{2}+i0^{+}}-\frac{\gamma ^{+}}{%
2p^{+}}, \
\end{equation}%
where the noncovariant second term on the right-hand side is present in the propagator of the non-dynamical component of the Dirac field, according to the canonical approach \cite{ref12}. The importance of that term has been discussed in many works. For instance, it has been shown that this term does not propagate any information \cite{ref50} and interferes in the preservation of the Ward identity \cite{ref51}. Also, by explicit calculation, it has been demonstrated that the contribution of such noncovariant term to the gluon self-energy \cite{ref12} and also to the electron self-energy \cite{ref14} are compensated in the context of the dimensional regularization scheme.

If we want to analyse the noncovariant term in the causal approach framework, without any modification of the splitting formula, we must consider the fermionic Green's function in the following form%
\begin{equation}
S_{I}\left( p\right) =\frac{\gamma .p+m}{p^{2}-m^{2}}-\frac{\gamma ^{+}}{%
2p^{+}},  \label{eq 3.5}
\end{equation}%
where the last term is named as the \emph{instantaneous} part, and this Green's functions is called incomplete. Before discussing the one-loop vacuum polarization version for the incomplete case \textit{per se}, we shall analyze the incomplete propagators associated to \eqref{eq 3.5} in the causal program in order to shed some new light in this long-term recognized problem.

\subsection{Fermionic propagator with instantaneous part}

It is well-known that the causal program has its beginning from the free field equation. Then, as a first step in discussing the incomplete case \eqref{eq 3.5}, we look for a field equation which reproduces the incomplete Green's function:%
\begin{equation}
S_{I}\left( p\right) =\frac{\gamma .p+m}{p^{2}-m^{2}}-\frac{\gamma ^{+}}{2p^{+}}.  \label{eq 3.6}
\end{equation}%
Now, considering that $\mathcal{D}\left( \partial \right) \psi =0$ is the free field equation, then by definition the Green's function is given by%
\begin{equation}
\mathcal{D}\left( p\right) =-\left[ S_{I}\left( p\right) \right] ^{-1}. \label{eq 3.7}
\end{equation}%
Moreover, if we denote the incomplete Green's function as $S_{I}\left(p,m\right) $, and use the properties:%
\begin{equation}
\left( \gamma ^{+}\right) ^{2}=0,\quad \left( \gamma .p\right) \gamma
^{+}+\gamma ^{+}\left( \gamma .p\right) =2p^{+},  \label{eq 3.8}
\end{equation}%
we can easily show that%
\begin{equation}
S_{I}\left( p,m\right) S_{I}\left( p,-m\right) =0.
\end{equation}%
Then, from the Binet's theorem: $\det \left[ AB\right] =\det \left[ A\right]\det \left[ B\right] $, we have%
\begin{equation}
\det \left[ S_{I}\left( p,m\right) \right] =0,\text{  or }\det \left[
S_{I}\left( p,-m\right) \right] =0.  \label{eq 3.9}
\end{equation}%
In general, looking only onto the field equation, we can consider $m$ as being an arbitrary real number, then
\begin{equation}
\det \left[ S_{I}\left( p,m\right) \right] =0.  \label{eq 3.10}
\end{equation}%
Therefore, it follows that $S_{I}\left( p\right) $ does not has an inverse. This means that must exists an \emph{additional constraint} to the Dirac equation in such a way that it generates the incomplete Green's function $S_{I}\left(p\right) $. Though we know that $S_{I}\left( p\right) $ has not direct relation to the Dirac equation, we can use the causal approach to determine the PF and NF propagators for this case \cite{ref28}:%
\begin{equation}
\left\langle \hat{S}_{I}^{\left( \pm \right) },\varphi \right\rangle =\left(2\pi \right) ^{-2}\theta \left( \pm p^{-}\right) \oint\limits_{c_{all}}S_{I}\left( p\right) \varphi \left( p^{+}\right) dp^{+},  \label{eq 3.11}
\end{equation}%
where $c_{all}$ are all counterclockwise closed paths which contain all individual poles in the complex plane of $p^{+}$. After some algebraic manipulation we arrive at
\begin{equation}
\hat{S}_{I}^{\left( \pm \right) }\left( p\right) =\left( \gamma .p+m\right)
\hat{D}_{m}^{\left( \pm \right) }\left( p\right) -\frac{\gamma ^{+}}{2}\hat{D%
}_{1}^{\left( \pm \right) }\left( p\right) ,  \label{eq 3.12}
\end{equation}%
where $\hat{D}_{1}^{\left( \pm \right) }\left( p\right) =\frac{i}{2\pi }\theta \left( \pm p^{-}\right) \delta \left( p^{+}\right) $. Then, the causal propagator is given by%
\begin{equation}
\hat{S}_{I}\left( p\right) =\hat{S}_{I}^{\left( +\right) }\left( p\right) +%
\hat{S}_{I}^{\left( -\right) }\left( p\right) =\left( \gamma .p+m\right)
\hat{D}_{m}\left( p\right) -\frac{i}{2\pi }\gamma ^{+}\delta \left(
p^{+}\right) .  \label{eq 3.13}
\end{equation}%
Nonetheless, it is interesting to rewrite \eqref{eq 3.13} in the configuration space. Thus, we obtain the fermionic causal propagator with the instantaneous part%
\begin{equation}
S_{I}\left( x\right) =S\left( x\right) -i\gamma ^{+}\delta \left(
x^{+}\right) \delta \left( x^{\bot }\right) .  \label{eq 3.15}
\end{equation}%
Since $S$ has causal support then $S_{I}$ has it as well, this means that we shall not have any problem with non-locality in \eqref{eq 3.15}.

\subsection{Vacuum polarization tensor at one-loop with instantaneous part}

From the results of Sec.\ref{sec:2}, we know that for the vacuum polarization contribution at one-loop we need to consider the distributions:
\begin{align}
R_{2}^{\prime VP}\left( x_{1},x_{2}\right) &= -\colon A_{\mu }\left( x_{1}\right) P^{\nu \mu }_{I}\left( x_{2}-x_{1}\right) A_{\nu }\left( x_{2}\right) \colon  \label{eq 3.16} \\
D_{2}^{VP}\left( x_{1},x_{2}\right) &=  \colon A_{\mu }\left( x_{1}\right) \left[ P_{I}^{\mu \nu }\left( x_{1}-x_{2}\right) -P_{I}^{\nu \mu }\left( x_{2}-x_{1}\right) \right] A_{\nu }\left( x_{2}\right) \colon ,
\label{eq 3.17}
\end{align}%
where $P_{I}^{\mu \nu }\left( y\right) =e^{2}tr\left[ \gamma ^{\mu }S_{I}^{\left( +\right) }\left( y\right) \gamma ^{\nu }S_{I}^{\left(-\right) }\left( -y\right) \right] $. Besides, as it was aforementioned, it is necessary to prove that $D_{2}$ or its numerical part $d^{\mu \nu }$,
\begin{equation}
d^{\mu \nu }\left( x_{1}-x_{2}\right) =P_{I}^{\mu \nu }\left( x_{1}-x_{2}\right)
-P_{I}^{\nu \mu }\left( x_{2}-x_{1}\right) ,
\end{equation}%
has causal support. We then can show that $d^{\mu \nu }$ can be written as the following%
\begin{equation}
e^{-2}d^{\mu \nu }\left( x_{1}-x_{2}\right) =tr\left\{ \gamma ^{\mu
}S_{I}^{\left( +\right) }\left( x_{1}-x_{2}\right) \gamma ^{\nu }S_{I}\left(
x_{2}-x_{1}\right) \right\} -tr\left\{ \gamma ^{\mu }S_{I}\left(
x_{1}-x_{2}\right) \gamma ^{\nu }S_{I}^{\left( +\right) }\left(
x_{2}-x_{1}\right) \right\} .  \label{eq 3.18}
\end{equation}%
From the last section results, we know that $S_{I}$ has causal support, so the product $\left( S_{I}^{\left( +\right) }S_{I}\right) $ has causal support as well. Hence, we may conclude that the distribution $d^{\mu \nu }$ has causal support as required. We should emphasize, however, that the only difference with those results from the Sec.\ref{sec:2} is in the tensor $P_{I}^{\mu \nu }$. Therefore, a suitable first step in the analysis it would be to show whether or not there is a difference between the expressions of this tensor and $P^{\mu \nu }$, Eq.\eqref{eq 2.12}.

In order to look up for the \emph{difference} between the quantities $P_{I}^{\mu \nu }$ and $P^{\mu \nu }$ we shall write down explicitly the tensor $P^{\mu\nu }$,%
\begin{equation}
\hat{P}^{\mu \nu }\left( k\right) =e^{2}\left( 2\pi \right) ^{-2}tr\int
d^{4}p\gamma ^{\mu }\hat{S}^{\left( +\right) }\left( p\right) \gamma ^{\nu }%
\hat{S}^{\left( -\right) }\left( p-k\right) ,  \label{eq 3.19}
\end{equation}%
in terms of the incomplete propagators, since by definition \eqref{eq 3.12} we have%
\begin{equation}
\hat{S}^{\left( \pm \right) }=\hat{S}_{I}^{\left( \pm \right) }+\frac{\gamma
^{+}}{2}\hat{D}_{1}^{\left( \pm \right) }.
\end{equation}%
From the complete expression, we can identify four different parts of $\hat{P}^{\mu \nu }\left( k\right) $:%
\begin{equation}
\hat{P}^{\mu \nu }\left( k\right) =\hat{P}_{I}^{\mu \nu }\left( k\right) +%
\hat{P}_{I1}^{\mu \nu }\left( k\right) +\hat{P}_{1I}^{\mu \nu }\left(
k\right) +\hat{P}_{11}^{\mu \nu }\left( k\right) ,  \label{eq 3.20}
\end{equation}%
where
\begin{subequations}
\begin{align}
\hat{P}_{I}^{\mu \nu }\left( k\right) &= e^{2}\left( 2\pi \right) ^{-2}\int d^{4}ptr\left[ \gamma ^{\mu }\hat{S}_{I}^{\left( +\right) }\left( p\right) \gamma ^{\nu }\hat{S}_{I}^{\left( -\right) }\left( p-k\right) \right] , \\
\hat{P}_{I1}^{\mu \nu }\left( k\right) &= e^{2}\left( 2\pi \right) ^{-2}\frac{1}{2}\int d^{4}ptr\left[ \gamma ^{\mu }\hat{S}_{I}^{\left( +\right) }\left( p\right) \gamma ^{\nu }\gamma ^{+}\hat{D}_{1}^{\left( -\right)
}\left( p-k\right) \right] , \\
\hat{P}_{1I}^{\mu \nu }\left( k\right) &= e^{2}\left( 2\pi \right) ^{-2}\frac{1}{2}\int d^{4}ptr\left[ \gamma ^{\mu }\hat{D}_{1}^{\left( +\right) }\left( p\right) \gamma ^{+}\gamma ^{\nu }\hat{S}_{I}^{\left( -\right)
}\left( p-k\right) \right] , \\
\hat{P}_{11}^{\mu \nu }\left( k\right) &= e^{2}\left( 2\pi \right) ^{-2}\frac{1}{4}\int d^{4}ptr\left[ \gamma ^{\mu }\gamma ^{+}\hat{D}_{1}^{\left(+\right) }\left( p\right) \gamma ^{\nu }\gamma ^{+}\hat{D}_{1}^{\left(
-\right) }\left( p-k\right) \right] .
\end{align}
\end{subequations}
After some calculation, see \ref{app:C}, we find, from the Eqs.\eqref{eq c.8} and \eqref{eq c.14}, the results:
\begin{equation}
\hat{P}_{11}^{\mu \nu }\left( k\right) =0,\quad \text{and}\quad \hat{P}%
_{1I}^{\mu \nu }\left( k\right) +\hat{P}_{I1}^{\mu \nu }\left( k\right) =0.\label{eq 3.21}
\end{equation}%
Therefore, we have shown that the original and incomplete tensors are in fact identical, Eq.\eqref{eq c.15}:%
\begin{equation}
\hat{P}^{\mu \nu }\left( k\right) =\hat{P}_{I}^{\mu \nu }\left( k\right) .\label{eq 3.22}
\end{equation}%
From this result we can conclude that all the results obtained in the Sec.\ref{sec:2} are valid to this incomplete case, in particular the vacuum polarization tensor:%
\begin{equation}
\hat{\Pi}_{I}^{\mu \nu }\left( k\right) =\hat{\Pi}^{\mu \nu }\left( k\right).  \label{eq 3.23}
\end{equation}%
We see then that this conclusion follows from general distributional theory requirements only, which is rather satisfactory from the point-of-view of the generality of the Epstein-Glaser causal framework.

\section{Concluding Remarks}

\label{sec:4}

In this paper we have implemented the causal perturbation theory of Epstein-Glaser to a field theory defined in the light-front form. This may be named as the Epstein-Glaser-Dirac causal method. This new approach to the \textit{S-Matrix}
in the light-front form gave us well-defined results, in the sense that they are finite and fulfill general physical requirements, such as causality, in each step of this perturbative program.

In this approach, we calculated the one-loop vacuum polarization for the light-front QED$_{4}$ in full detail. For this calculation we had considered the light-front transversality in the vacuum polarization tensor. On the other hand, in order to obtain the final expression for the vacuum polarization tensor we applied the polynomial normalization of the perturbative causal method, which is somehow similar to the usual renormalization program, but without having regularized divergent integrals, and we have found that our main central splitting solution fulfill the physical considerations: parity, photon mass shell and charge normalization.

Finally, we analyzed the case of the instantaneous part of the fermionic propagator. We showed that this case can not be considered in the causal method, by relying on the simple argument: that unless including further constraints to the Dirac equation we may not obtain a propagator such as \eqref{eq 3.12}. But, if we consider this part in the calculation, nevertheless, we may show that it does not contributes to the vacuum polarization expression. This result is so transparent and clear here, because the calculations are taken in part of the distributional theory, the mathematical framework of the Epstein-Glaser causal approach.

The full strength of the causal method of Epstein-Glaser has been exploited in many studies in the framework of field theoretical models along the years, and in light of that strength we have decided to use the method also in the light-front field theories, which stand nowadays in our opinion as one of the most richest frameworks to be studied; but, at the same time, it is plagued with dubious and ill-defined formal issues. So, we have made use previously of the causal theory to discuss free fields in light-front, in particular, discussing the behavior of the light-front singularity of the type $g\left(k;n\right)=1/\left( k^{+}\right) ^{n}$, and now in the present paper we have showed how powerful the causal approach may also be in dealing with interacting fields in the light-front. There are many interesting related issues within light-front field theories that deserve to be analyzed carefully, especially in the light-front QCD$_{4}$ \cite{ref71}, where many efforts have been applied either in the perturbative and nonperturbative regime, and some others issues in different context that we believe that the causal theory may shed some new and fresh light to some inherent illness, in such a way to obtain well-defined and unambiguous outcomes. These issues and others will be further elaborated, investigated and reported elsewhere.

\subsection*{Acknowledgments}

R.B. thanks FAPESP for full support, B.M.P. thanks CNPq and CAPES for partial support and D.E.S. thanks CNPq for full support.

\appendix

\section{Light-front notation and properties}

\label{app:A}

If $\left( x^{0},x^{1},x^{2},x^{3}\right) $ are the instant-form coordinates, then the light-front coordinates, $\left(
x^{+},x^{1},x^{2},x^{-}\right) $, are related to these by the following
relations%
\begin{equation}
\hat{x}^{0,3} =\frac{x^{0}\pm x^{3}}{\sqrt{2}}\equiv x^{\pm }, \quad
\hat{x}^{\bot } =\left( x^{1},x^{2}\right) . \label{eq a.1}
\end{equation}%
Moreover, the metric in the light-front form is given by%
\begin{equation}
h_{\mu \nu }=h^{\mu \nu }=\allowbreak
\begin{pmatrix}
0 & 0 & 0 & 1 \\
0 & -1 & 0 & 0 \\
0 & 0 & -1 & 0 \\
1 & 0 & 0 & 0%
\end{pmatrix}%
.
\end{equation}%
The invariant inner product takes the form%
\begin{equation}
A_{\mu }B^{\mu }=2A_{+}B_{-}-A_{\bot }.B_{\bot },
\end{equation}%
where the components of the vector $A^{\mu }=\left( A^{+},A^{\bot},A^{-}\right) $ are usually denoted as the temporal, transversal and longitudinal components, respectively.

Similar definitions are applied to the Dirac matrices that still obey the anticommutation relation%
\begin{equation}
\left\{ \gamma ^{\mu },\gamma ^{\nu }\right\} =2h^{\mu \nu }.  \label{eq a.2}
\end{equation}%
In particular we have that: $\left( \gamma ^{+}\right) ^{2}=\left( \gamma ^{-}\right) ^{2}=0$. Moreover, they satisfy similar trace properties as the instant form%
\begin{align}
tr\left( \gamma ^{\mu }\gamma ^{\nu }\right) &= 4h^{\mu \nu }, \\
tr\left( \gamma ^{\mu }\gamma ^{\alpha }\gamma ^{\nu }\gamma ^{\beta
}\right) &= 4\left( h^{\mu \alpha }h^{\nu \beta }-h^{\mu \nu }h^{\alpha
\beta }+h^{\mu \beta }h^{\alpha \nu }\right) ,\\
tr\left( \gamma ^{\mu }\cdots \gamma ^{\mu _{2n-1}}\right) &= 0,\quad \forall ~ n\in \mathbb{N}.
\end{align}

\section{Convolution in the Light-front}

\label{app:B}

In this appendix we calculate in details the convolution $\hat{D}_{m}^{\left( +\right) }\ast \hat{D}_{m}^{\left( -\right) }$, which appeared in the Eq.\eqref{eq 2.19}. We have that it is defined as follows:
\begin{equation}
f\left( k\right) =\int d^{4}p\hat{D}_{m}^{\left( +\right) }\left( p\right)
\hat{D}_{m}^{\left( -\right) }\left( p-k\right) ,  \label{eq b.1}
\end{equation}%
where $\hat{D}_{m}^{\left( \pm \right) }\left( p\right) =\pm \frac{i}{2\pi}\delta \left( 2p^{+}p^{-}-\omega_{m} ^{2}\right) \theta \left( \pm p^{-}\right)$, and $\omega _{m}^{2}=p_{\bot }^{2}+m^{2}\geq m^{2}$. Now, replacing the scalar propagators \cite{ref28}:%
\begin{align}
f\left( k\right) &= \left( 2\pi \right) ^{-2}\int d^{4}p\theta \left(
p^{-}\right) \theta \left( k^{-}-p^{-}\right) \delta \left(
2p^{+}p^{-}-\omega _{m}^{2}\right)  \notag \\
& \quad \times \delta \left( -2k^{+}p^{-}+2k^{+}k^{-}-2p^{+}k^{-}+2p^{+}p^{-}-\bar{%
\omega}_{m}^{2}\right) ,  \label{eq b.2}
\end{align}%
in which $\bar{\omega}_{m}^{2}=\left(p_{\bot }-k_{\bot }\right) ^{2}+m^{2}$. We may calculate conveniently this integral in a referential system such that $k=\left( k^{+},k^{-}\right)\equiv \left( k^{+},0^{\bot },k^{-}\right) $, so $\bar{\omega}_{m}^{2}=\omega _{m}^{2}$, then
\begin{align}
f\left( k^{+},k^{-}\right) &= \left( 2\pi \right) ^{-2}\int d^{4}p\theta
\left( p^{-}\right) \theta \left( k^{-}-p^{-}\right) \delta \left(2p^{+}p^{-}-\omega _{m}^{2}\right)  \notag \\
& \quad \times \delta \left(-2k^{+}p^{-}+2k^{+}k^{-}-2p^{+}k^{-}+2p^{+}p^{-}-\omega _{m}^{2}\right) .
\label{eq b.3}
\end{align}%
From the very definition of the $\delta $-Dirac and using the scaling property $\delta \left( \alpha x\right) =\frac{1}{\left\vert \alpha \right\vert }\delta \left( x\right) $, we may integrate in $p^{+}$, resulting into:%
\begin{equation}
f\left( k^{+},k^{-}\right) =\left( 2\pi \right) ^{-2}\int d^{2}p^{\bot
}dp^{-} \frac{\theta \left( p^{-}\right) \theta \left( k^{-}-p^{-}\right)}{%
\left\vert 4k^{+}\right\vert }\delta \left[ \left( p^{-}\right)
^{2}-p^{-}k^{+}k^{-}-\frac{k^{-}}{2k^{+}}\omega _{m}^{2}\right] .
\label{eq b.4}
\end{equation}%
By a boost transformation it is possible to have $k^{+}=k^{-}$,%
\begin{equation}
f\left( k^{-},k^{-}\right) =\frac{\left( 2\pi \right) ^{-2}}{4\left\vert
k^{-}\right\vert }\int d^{2}p^{\bot }dp^{-}\theta \left( p^{-}\right)
\theta \left( k^{-}-p^{-}\right) \delta \left[ \left( p^{-}-\frac{k^{-}}{2}%
\right) ^{2}- \frac{\left( k^{-}\right) ^{2}}{4}+\frac{\omega _{m}^{2}%
}{2} \right] .  \label{eq b.5}
\end{equation}%
By spherical symmetry of the integral in respect to $p^{\bot }$, so: $d^{2}p^{\bot }=\pi d\omega _{m}^{2}$, we obtain%
\begin{equation}
f\left( k^{-},k^{-}\right) =\frac{\left( 2\pi \right) ^{-1}}{8\left\vert k^{-}\right\vert }\int d\omega _{m}^{2}dp^{-}\theta \left( p^{-}\right) \theta \left( k^{-}-p^{-}\right) \delta \left[ \left( p^{-}-\frac{k^{-}}{2}%
\right) ^{2}- \frac{\left( k^{-}\right) ^{2}}{4}+\frac{\omega _{m}^{2}}{2} \right] .  \label{eq b.6}
\end{equation}%
From the argument of the $\delta $-distribution we find that the non-null case is given in the region%
\begin{equation}
m^{2}\leq \omega _{m}^{2}\leq \frac{\left( k^{-}\right) ^{2}}{2},\quad \text{%
and}\quad 0\leq p^{-}\leq k^{-}.
\end{equation}%
We can work the expression \eqref{eq b.6} by means of a property of the $\delta $-Dirac distribution, $a= p^{-}-\frac{k^{-}}{2}$ and $b=\sqrt{\frac{\left( k^{-}\right) ^{2}}{4}-\frac{\omega ^{2}_m}{2}}$ we have that%
\begin{align}
\delta \left[a^2-b^2\right]=\frac{1}{|2b|}\left\lbrace\delta \left[a-b\right]+\delta \left[a+b\right]\right\rbrace.\label{eq b.7}
\end{align}%
Therefore, in the expression \eqref{eq b.6} we have $0\leq p^{-}\leq k^{-}$ and since $0\leq \frac{k^{-}}{2}\pm \sqrt{\frac{\left( k^{-}\right) ^{2}}{4}-\frac{\omega ^{2}}{2}}\leq k^{-}$, one may integrate both in $p^{-}$ and $%
\omega _{m}^{2}$, to finally obtain that:%
\begin{equation}
f\left( k^{-},k^{-}\right) =\left( 2\pi \right) ^{-1}\frac{1}{4}\theta
\left( k^{-}\right) \theta \left[ 2\left( k^{-}\right) ^{2}-4m^{2}\right]
\sqrt{1-\frac{4m^{2}}{2\left( k^{-}\right) ^{2}}}.  \label{eq b.8}
\end{equation}%
We may return, by a boost and rotation, to an arbitrary referential system, we then have that for an arbitrary $k\in \Gamma ^{+\text{:}}$%
\begin{equation}
f\left( k\right) =\left( 2\pi \right) ^{-1}\frac{1}{4}\theta \left(
k^{-}\right) \theta \left[ k^{2}-4m^{2}\right] \sqrt{1-\frac{4m^{2}}{k^{2}}}.
\label{eq b.9}
\end{equation}%
This result is very similar to the one in usual coordinates, which can be obtained by the change: $k^{-}\rightarrow k^{0}$.

\section{Decomposition of the tensor $P^{\protect\mu \protect\nu }$}

\label{app:C}

In analyzing the relation between the original $\hat{P}^{\mu \nu }\left(k\right) $ and incomplete $\hat{P}_{II}^{\mu \nu }\left( k\right) $ tensors we obtained the following expression $\hat{P}^{\mu \nu }\left( k\right) $ \eqref{eq 3.20}%
\begin{equation}
\hat{P}^{\mu \nu }\left( k\right) =\hat{P}_{I}^{\mu \nu }\left( k\right) +%
\hat{P}_{I1}^{\mu \nu }\left( k\right) +\hat{P}_{1I}^{\mu \nu }\left(
k\right) +\hat{P}_{11}^{\mu \nu }\left( k\right) ,  \label{eq c.1}
\end{equation}%
in which
\begin{subequations}
\begin{align}
\hat{P}_{I}^{\mu \nu }\left( k\right) &= e^{2}\left( 2\pi \right) ^{-2}\int
d^{4}ptr\left[ \gamma ^{\mu }\hat{S}_{I}^{\left( +\right) }\left( p\right)
\gamma ^{\nu }\hat{S}_{I}^{\left( -\right) }\left( p-k\right) \right] ,
\label{eq c.2} \\
\hat{P}_{I1}^{\mu \nu }\left( k\right) &= e^{2}\left( 2\pi \right) ^{-2}%
\frac{1}{2}\int d^{4}ptr\left[ \gamma ^{\mu }\hat{S}_{I}^{\left( +\right)
}\left( p\right) \gamma ^{\nu }\gamma ^{+}\hat{D}_{1}^{\left( -\right)
}\left( p-k\right) \right] ,  \label{eq c.3} \\
\hat{P}_{1I}^{\mu \nu }\left( k\right) &= e^{2}\left( 2\pi \right) ^{-2}%
\frac{1}{2}\int d^{4}ptr\left[ \gamma ^{\mu }\hat{D}_{1}^{\left( +\right)
}\left( p\right) \gamma ^{+}\gamma ^{\nu }\hat{S}_{I}^{\left( -\right)
}\left( p-k\right) \right] ,  \label{eq c.4} \\
\hat{P}_{11}^{\mu \nu }\left( k\right) &= e^{2}\left( 2\pi \right) ^{-2}%
\frac{1}{4}\int d^{4}ptr\left[ \gamma ^{\mu }\gamma ^{+}\hat{D}_{1}^{\left(
+\right) }\left( p\right) \gamma ^{\nu }\gamma ^{+}\hat{D}_{1}^{\left(
-\right) }\left( p-k\right) \right] .  \label{eq c.5}
\end{align}%
\end{subequations}
We shall now proceed in evaluating separately each one of these parts:

\emph{Instantaneous} part, by definition $\hat{P}_{11}^{\mu \nu }$ is given by
\begin{equation}
\hat{P}_{11}^{\mu \nu }\left( k\right) =e^{2}\left( 2\pi \right) ^{-2}\frac{1%
}{4}tr\left( \gamma ^{\mu }\gamma ^{+}\gamma ^{\nu }\gamma ^{+}\right) \int
d^{4}p\hat{D}_{1}^{\left( +\right) }\left( p\right) \hat{D}_{1}^{\left(
-\right) }\left( p-k\right) .  \label{eq c.6}
\end{equation}%
We shall evaluate first the momentum integral%
\begin{equation}
f_{11}\left( k\right) =\int d^{4}p\hat{D}_{1}^{\left( +\right) }\left(
p\right) \hat{D}_{1}^{\left( -\right) }\left( p-k\right) .
\end{equation}%
Since $\hat{D}_{1}^{\left( \pm \right) }\left( q\right) =\frac{i}{2\pi }%
\theta \left( \pm q^{-}\right) \delta \left( q^{+}\right) $, it follows%
\begin{align}
f_{11}\left( k\right) &= -\left( 2\pi \right) ^{-2}\int d^{4}p\theta \left(
p^{-}\right) \theta \left( k^{-}-p^{-}\right) \delta \left( p^{+}\right)
\delta \left( p^{+}-k^{+}\right) , \notag \\
&= -\left( 2\pi \right) ^{-2}\theta \left( k^{-}\right) \int d^{2}p^{\bot
}dp^{+}\left( k^{-}\right) \delta \left( p^{+}\right) \delta \left(
p^{+}-k^{-}\right) ,  \label{eq c.7}
\end{align}%
in which we are considering a referential frame such that $k=\left(k^{-},0^{\bot },k^{-}\right) $. This expression can also be rewritten as it follows%
\begin{align}
f_{11}\left( k\right) =&-\left( 2\pi \right) ^{-2}\theta \left(
k^{-}\right) \int d^{2}p^{\bot }\int dp^{+}\left( k^{-}-p^{+}\right) \delta
\left( p^{+}-k^{-}\right) \delta \left( p^{+}\right)  \notag \\
&-\left( 2\pi \right) ^{-2}\theta \left( k^{-}\right) \int d^{2}p^{\bot
}\int dp^{+}\left( p^{+}\right) \delta \left( p^{+}\right) \delta \left(
p^{+}-k^{-}\right) .
\end{align}%
Finally, one may show, by using the property $x\delta \left( x\right) =0$, that $f_{11}\left( k\right) =0$. Therefore, it follows the result
\begin{equation}
\hat{P}_{11}^{\mu \nu }\left( k\right) =0.  \label{eq c.8}
\end{equation}%

\emph{Mixed} parts, for this case we have two contributions. We shall consider first:%
\begin{equation}
\hat{P}_{I1}^{\mu \nu }\left( k\right) =e^{2}\left( 2\pi \right) ^{-2}\frac{1}{2}tr\int d^{4}p\gamma ^{\mu }\hat{S}_{I}^{\left( +\right) }\left( p\right) \gamma ^{\nu }\gamma ^{+}\hat{D}_{1}^{\left( -\right) }\left( p-k\right) ,
\label{eq c.9}
\end{equation}%
Moreover, since $\hat{S}_{I}^{\left( +\right) }\left( p\right) =\hat{S}^{\left( +\right) }\left( p\right) -\frac{\gamma ^{+}}{2}\hat{D}_{1}^{\left( +\right) }$, it follows%
\begin{align}
\hat{P}_{I1}^{\mu \nu }\left( k\right) &= e^{2}\left( 2\pi \right) ^{-2}%
\frac{1}{2}tr\int d^{4}p\gamma ^{\mu }\hat{S}^{\left( +\right) }\left(
p\right) \gamma ^{\nu }\gamma ^{+}\hat{D}_{1}^{\left( -\right) }\left(
p-k\right)  \notag \\
& \quad -e^{2}\left( 2\pi \right) ^{-2}\frac{1}{4}tr\int d^{4}p\gamma ^{\mu
}\gamma ^{+}\hat{D}_{1}^{\left( +\right) }\left( p\right) \gamma ^{\nu
}\gamma ^{+}\hat{D}_{1}^{\left( -\right) }\left( p-k\right) .
\end{align}%
From that, we can identify the last term as $-\hat{P}_{11}^{\mu \nu }\left(k\right) $ , so this term does not contribute. Next, replacing the definition: $\hat{S}^{\left( +\right) }=\left( \gamma .p+m\right) \hat{D}_{m}^{\left( +\right) }\left( p\right) $ and, using some properties of the $\gamma $-matrices, we obtain for the first term
\begin{equation}
\hat{P}_{I1}^{\mu \nu }\left( k\right) =e^{2}\left( 2\pi \right) ^{-2}\frac{1%
}{2}\int d^{4}ptr\left( \gamma ^{\mu }\left( \gamma .p\right) \gamma ^{\nu
}\gamma ^{+}\right) \hat{D}_{m}^{\left( +\right) }\left( p\right) \hat{D}%
_{1}^{\left( -\right) }\left( p-k\right) .  \label{eq c.10}
\end{equation}%
We consider now the second mixed part $\hat{P}_{1I}^{\mu \nu }$ which is defined as%
\begin{equation}
\hat{P}_{1I}^{\mu \nu }\left( k\right) =e^{2}\left( 2\pi \right) ^{-2}\frac{1%
}{2}tr\int d^{4}p\gamma ^{\mu }\hat{D}_{1}^{\left( +\right) }\left( p\right)
\gamma ^{+}\gamma ^{\nu }\hat{S}_{I}^{\left( -\right) }\left( p-k\right) .
\label{eq c.11}
\end{equation}%
Since $\hat{S}_{I}^{\left( -\right) }\left( p\right) =\hat{S}^{\left(-\right) }\left( p\right) -\frac{\gamma ^{+}}{2}\hat{D}_{1}^{\left( -\right)}$ and $\hat{S}^{\left( -\right) }\left( p\right) =\left( \gamma .p+m\right)
\hat{D}_{m}^{\left( -\right) }\left( p\right) $, one may follows the steps as outlined in the previous case, and obtain that
\begin{equation}
\hat{P}_{1I}^{\mu \nu }\left( k\right) =e^{2}\left( 2\pi \right) ^{-2}\frac{1%
}{2}\int d^{4}ptr\left( \gamma ^{\mu }\gamma ^{+}\gamma ^{\nu }\left( \gamma
.p-\gamma .k\right) \right) \hat{D}_{1}^{\left( +\right) }\left( p\right)
\hat{D}_{m}^{\left( -\right) }\left( p-k\right) .
\end{equation}%
Making the variables change: $q=k-p$, we have%
\begin{equation}
\hat{P}_{1I}^{\mu \nu }\left( k\right) =e^{2}\left( 2\pi \right) ^{-2}\frac{1%
}{2}\int d^{4}qtr\left( \gamma ^{\mu }\gamma ^{+}\gamma ^{\nu }\gamma
.q\right) \hat{D}_{m}^{\left( -\right) }\left( -q\right) \hat{D}_{1}^{\left(
+\right) }\left( k-q\right) .  \label{eq c.12}
\end{equation}%
By the trace property: $tr\left( \gamma ^{\mu }\gamma ^{+}\gamma ^{\nu }\gamma .q\right) =tr\left( \gamma ^{\mu }\gamma .q\gamma ^{\nu }\gamma ^{+}\right) $, and the following relations of the PF and NF propagators
\begin{align}
\hat{D}_{m}^{\left( -\right) }\left( -q\right) &= -\hat{D}_{m}^{\left(+\right) }\left( q\right), \\
\hat{D}_{1}^{\left( +\right) }\left( k-q\right) &= \frac{i}{2\pi }\theta %
\left[ -\left( q^{-}-k^{-}\right) \right] \delta \left( q^{+}-k^{+}\right) =%
\hat{D}_{1}^{\left( -\right) }\left( q-k\right) ,
\end{align}%
we arrive at%
\begin{equation}
\hat{P}_{1I}^{\mu \nu }\left( k\right) =-e^{2}\left( 2\pi \right) ^{-2}\frac{%
1}{2}\int d^{4}qtr\left( \gamma ^{\mu }\gamma .q\gamma ^{\nu }\gamma
^{+}\right) \hat{D}_{m}^{\left( +\right) }\left( q\right) \hat{D}%
_{1}^{\left( -\right) }\left( q-k\right) .  \label{eq c.13}
\end{equation}%
By comparing this result with \eqref{eq c.10}, we may conclude that
\begin{equation}
\hat{P}_{1I}^{\mu \nu }\left( k\right) =-\hat{P}_{I1}^{\mu \nu }\left(
k\right) .  \label{eq c.14}
\end{equation}%
Therefore, the relation \eqref{eq c.1} reads%
\begin{equation}
\hat{P}^{\mu \nu }\left( k\right) =\hat{P}_{I}^{\mu \nu }\left( k\right) .
\label{eq c.15}
\end{equation}


\end{document}